\documentclass[11pt,a4paper]{article}
\usepackage[utf8]{inputenc}
\usepackage{tabularx}
\usepackage[dvipdfmx]{graphicx}
\usepackage[usenames]{xcolor}
\usepackage{colortbl}
\usepackage{hyperref}
\usepackage{siunitx}
\usepackage[top=30truemm,bottom=30truemm,left=31truemm,right=31truemm]{geometry}
\usepackage{setspace}
\usepackage{doi}
\usepackage{fancyhdr}
\usepackage{fancybox}
\usepackage{setspace}
\usepackage{amsmath}
\usepackage{wrapfig}
\usepackage{cmbright}
\usepackage[T1]{fontenc}

\usepackage{tikz}
\usetikzlibrary{trees}
\usetikzlibrary{decorations.pathmorphing}
\usetikzlibrary{decorations.markings}
\usetikzlibrary{decorations.pathreplacing}

\tikzset{
    photon/.style={decorate, decoration={snake},segment length=5.5pt, draw=black},
    fermion/.style={draw=black, postaction={decorate},
        decoration={markings,mark=at position .55 with {\arrow[draw=black]{>}}}},
    gluon/.style={decorate, draw=black,
        decoration={coil,amplitude=2.5pt, segment length=3.5pt}} 
}

\title{
\large Experiment Proposal \vspace{\baselineskip} \\
\Large Study of tau-neutrino production at the CERN SPS}
\author{
\small S. Aoki$^1$, A. Ariga$^2$, T. Ariga$^{2,3,}$\footnote{Contact person. E-mail: ariga@artsci.kyushu-u.ac.jp} , E. Firu$^4$, T. Fukuda$^5$, \\
\small Y. Gornushkin$^6$, A. M. Guler$^7$, M. Haiduc$^4$, K. Kodama$^8$, \\
\small M. A. Korkmaz$^7$, U. Kose$^9$, M. Nakamura$^5$, T. Nakano$^5$, \\
\small A. T. Neagu$^4$, H. Rokujo$^5$, O. Sato$^5$, S. Vasina$^6$, \\
\small M. Vladymyrov$^2$, M. Yoshimoto$^{10}$ \vspace{\baselineskip} \\
\fontsize{8pt}{0cm}\selectfont{$^1$}Kobe University, Kobe, Japan \\
\fontsize{8pt}{0cm}\selectfont{$^2$}AEC/LHEP, University of Bern, Bern, Switzerland \\
\fontsize{8pt}{0cm}\selectfont{$^3$}Kyushu University, Fukuoka, Japan \\
\fontsize{8pt}{0cm}\selectfont{$^4$}Institute of Space Science, Bucharest, Romania \\
\fontsize{8pt}{0cm}\selectfont{$^5$}Nagoya University, Nagoya, Japan \\
\fontsize{8pt}{0cm}\selectfont{$^6$}JINR-Joint Institute for Nuclear Research, Dubna, Russia \\
\fontsize{8pt}{0cm}\selectfont{$^7$}METU-Middle East Technical University, Ankara, Turkey \\
\fontsize{8pt}{0cm}\selectfont{$^8$}Aichi University of Education, Aichi, Japan \\
\fontsize{8pt}{0cm}\selectfont{$^9$}European Organization for Nuclear Research, Switzerland \\
\fontsize{8pt}{0cm}\selectfont{$^{10}$}Gifu University, Gifu, Japan \\ 
}
\date{}
\begin{document}
\maketitle

\newpage
\begin{abstract}
The DsTau project proposes to study tau-neutrino production in high-energy proton interactions. The outcome of this experiment are prerequisite for measuring the $\nu_\tau$ charged-current cross section that has never been well measured. Precisely measuring the cross section would enable testing of lepton universality in $\nu_\tau$ scattering and it also has practical implications for neutrino oscillation experiments and high-energy astrophysical $\nu_\tau$ observations. $D_s$ mesons, the source of tau neutrinos, following high-energy proton interactions will be studied by a novel approach to detect the double-kink topology of the decays $D_s \rightarrow \tau\nu_\tau$ and $\tau\rightarrow\nu_\tau X$. Directly measuring $D_s\rightarrow \tau$ decays will provide an inclusive measurement of the $D_s$ production rate and decay branching ratio to $\tau$. The momentum reconstruction of $D_s$ will be performed by combining topological variables. This project aims to detect 1,000 $D_s \rightarrow \tau$ decays in $2.3 \times 10^8$ proton interactions in tungsten target to study the differential production cross section of $D_s$ mesons. To achieve this, state-of-the-art emulsion detectors with a nanometric-precision readout will be used. The data generated by this project will enable the $\nu_\tau$ cross section from DONUT to be re-evaluated, and this should significantly reduce the total systematic uncertainty. Furthermore, these results will provide essential data for future $\nu_\tau$ experiments such as the $\nu_\tau$ program in the SHiP project at CERN. In addition, the analysis of $2.3 \times 10^8$ proton interactions, combined with the expected high yield of $10^5$ charmed decays as by-products, will enable the extraction of additional physical quantities.
\end{abstract}

\newpage
\tableofcontents

\newpage
\section{Physics motivations}
\subsection{Tau-neutrino production}

The DsTau project aims to study tau-neutrino production in high-energy proton interactions. The results of this experiment are prerequisite for measuring the $\nu_\tau$ charged-current (CC) cross section. Precisely measuring this will determine if lepton universality is applicable to $\nu_\tau$ scattering. 

Lepton universality is a basic assumption in the Standard Model (SM) of particle physics. 
A high precision measurement of neutrino interactions would be a powerful tool for testing lepton universality, whose violation would provide evidence for new physics effects. 
The decay branching ratios of a W boson to a charged lepton and neutrino have been measured at LEP and found to be equal within 2\% of the levels for electron, muon and tau \cite{lep}. However, lepton universality in $\nu_\tau$ scattering has never been well tested. 
Recent results of collider experiments (BABAR \cite{babar} and LHCb \cite{lhcb_b+,lhcb_b0}) indicate possible non-universality in leptonic decays of $B$ mesons. For example, $\bar{B} \rightarrow D^{(*)} \tau^- \bar{\nu}_\tau$ by BABAR \cite{babar} and $\bar{B}^0\rightarrow D^{*+}l^-\bar{\nu}_l$ ($l=\mu,\tau$) by LHCb \cite{lhcb_b0} give expressions 3.4 and 2.1 $\sigma$ away from the predictions of the SM, respectively, and the decay-branching ratios to $\tau\nu_\tau$ with respect to $\mu\nu_\mu$ are 25\%-35\% higher than the SM expectations. There are theoretical suggestions of lepton non-universality, particularly for $\nu_\tau$ \cite{nonuniversality_nutau}. Hence, experimental efforts in several directions are required to test for universality.

\begin{figure}[htbp]
\begin{center}
\includegraphics[width=9.6cm]{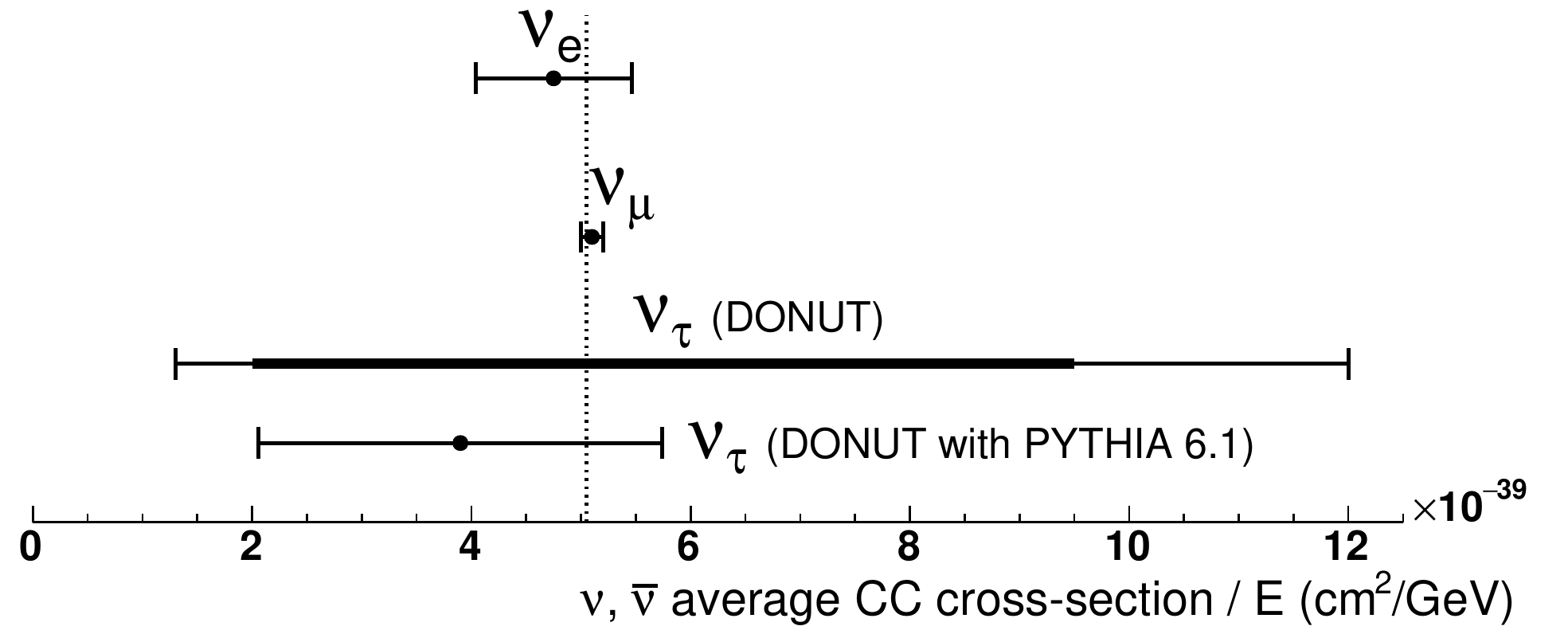}
\caption{$\nu$, $\overline{\nu}$ averaged energy independent cross-section of three neutrinos (\cite{gargamelle} for $\nu_e$, \cite{pdg2014} for $\nu_\mu$, and \cite{donut} for $\nu_\tau$). The vertical dashed line shows the SM prediction. 
}
\label{cross-sec-nu}
\end{center}
\end{figure}

Neutrino scattering, wherein the neutrino CC interaction cross sections are measured, can be used to study universality in the interactions of leptons. 
Any evidence for new physics is expected to appear in the $\tau$ sector. The $\nu_\tau$ CC cross section has not been properly studied until now. Figure \ref{cross-sec-nu} shows the measured energy-independent neutrino cross sections and their errors (an average of $\nu$ and $\overline{\nu}$). The SM prediction is identical for all three neutrinos and is indicated as a dashed vertical line. To date, there has only been one measurement for $\nu_\tau$ from the DONUT experiment \cite{donut}. 

The only practical way, as used in DONUT, of producing a $\nu_\tau$ beam is by the decay of charmed mesons, namely $D_s$. 
High-energy protons are sent to a target, creating $D_s$ mesons. These then decay in a chain, $D_s \rightarrow \tau\nu_\tau,\ \tau\rightarrow\nu_\tau X$, producing $\nu_\tau$ and $\overline{\nu}_\tau$ with every $D_s\rightarrow \tau$ decay. 
The main source of error in measuring DONUT's $\nu_\tau$ cross section is a result of systematic uncertainties, whereas 33\% of the relative uncertainty is due to the limited number of detected $\nu_\tau$ events (nine in total). The systematic uncertainty is much larger than 50\% and comes from the $\nu_\tau$ flux prediction. Indeed, owing to the lack of accurate measurements of the $D_s$ differential production cross section, the DONUT result was not a single value but a $\nu_\tau$ cross section as a function of a parameter $n$, which is responsible for the differential production cross section of $D_s$, as $\sigma^\textnormal{const}_{\nu_\tau}=2.51n^{1.52}\times 10^{-40}$ \si{cm^2} \si{GeV^{-1}}.
Therefore, the central value of the $\nu_\tau$ cross section was not defined (as shown in Figure \ref{cross-sec-nu}). The cross section can be calculated only if DONUT used a parameter value derived from PYTHIA 6.1 \cite{pythia61} simulations (as shown in Figure \ref{cross-sec-nu}). Note that this parameter value is significantly different in the recent version of PYTHIA. Since the non-universality effect can be in the range of 20\%-40\%, reducing the error of the $\nu_\tau$ cross section from its current value (>50\%) to 10\% would be a valuable result.

In the situation described above, it is vital to measure the differential production cross section of $D_s$ in high-energy proton interactions. It was rarely measured in previous experiments. The differential production cross-section of $D_s$ was conventionally approximated by a phenomenological formula, $d^2\sigma/(dx_Fdp^2_T)\propto (1-|x_F|)^n \exp(-bp^2_T)$, where $x_F$ is the Feynman $x$ ($x_F=2p^{CM}_Z/\sqrt{s}$) and $p_T$ is the transverse momentum. The first (second) term is a longitudinal (transverse) dependence controlled by a parameter $n$ ($b$). These parameters are to be obtained experimentally. The HERA-B experiment \cite{Hera-B}, which used a proton-beam energy of 920 GeV, reported 18.5 $\pm$ 7.6 $\mu$b/nucleon using 11.4 $\pm$ 4.0 $D_s^+$ events. The Fermilab experiments E653 \cite{E653} and E743 (LEBC-MPS) \cite{E743} measured the $D$ meson production in 800 GeV/c proton interactions; the measured transverse dependence $b$ ($D^0$, $D^+$) = 0.84 + 0.10 - 0.08 and $b$ ($D$) = 0.8 $\pm$ 0.2 were used in the DONUT analysis because the transverse dependence of charmed hadron production can be assumed to be the same for all charmed particles. The longitudinal dependence $n$ is known to be sensitive to the quark content as well as beam energy. E769 \cite{E769}, which used a 250 GeV proton beam, reported $n$ = 6.1 $\pm$ 0.7 for the inclusive charm meson production ($D^\pm$, $D^0$, $D_s^\pm$) \cite{E769}. All the other experiments using high-energy proton beams did not distinguish $D_s$ from all the other charmed particles ($D^\pm$, $D^0$, $\Lambda_c$) and only provided average values of $n$. Because the $D_s$ produced does not contain valence quarks from the initial proton, the leading particle effect is reduced. Thus, the differential production cross section should be different from other charmed particles, which could contain quarks of incoming protons. A dedicated measurement for $D_s$ is needed. E781 (SELEX) \cite{SELEX} used a 600-GeV $\Sigma^-$ beam (instead of a proton beam) and studied the difference between $D_s^+$ and $D_s^-$. The reported value for $D_s^+$, which is not affected by the leading particle effect, is $n$ = 7.4 $\pm$ 1.0 using about 130 $D_s^+$ events (within $x_F$ $>$ 0.15) in $\Sigma^-$ interactions (the value for $D_s^-$, which is affected by the leading particle effect, is $n$ = 4.1 $\pm$ 0.3). $D_s^+$ in SELEX may be similar to the DONUT situation; however, the incident particle is different, the measurement is limited to $x_F$ $>$ 0.15, and the uncertainty of $n$ is large owing to the limited data. These results are summarised in Table \ref{table_charm}. Results from the LHC experiments at $\sqrt{s}$ = 7, 8 or 13 TeV are not included here since the energies differ too much (400 GeV fixed target is at $\sqrt{s}$ = 27 GeV). In summary, no experimental result effectively constraining the $D_s$ differential cross section at the desired level or consequently the $\nu_\tau$ production exists.

\begin{table}[t]
\includegraphics[width=\linewidth]{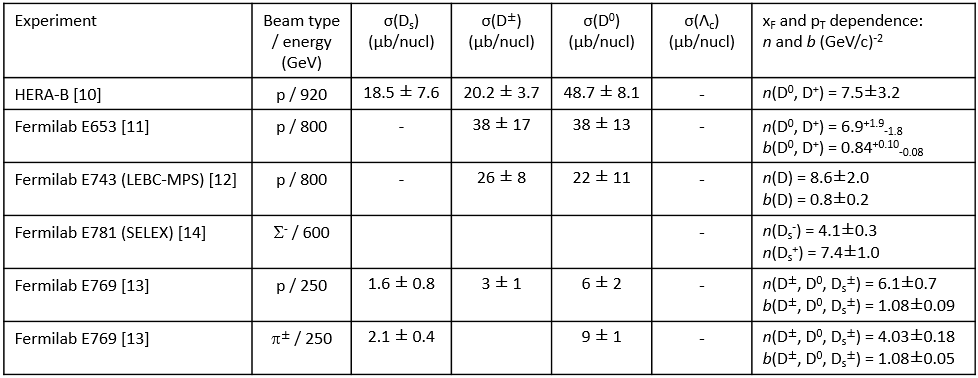}
\caption{Charm differential production cross section results obtained by the fixed-target experiments.}
\label{table_charm}
\end{table}

Here we propose a new approach to study $\nu_\tau$ production by measuring the $D_s\rightarrow \tau$ decays following high-energy proton interactions. This direct measurement of $D_s\rightarrow \tau$ decays will provide an inclusive measurement of the $D_s$ production and decay branching ratio to $\tau$ ($Br(D_s\rightarrow\tau) = (5.48 \pm 0.23)\%$ \cite{pdg2017}).
This project aims to detect 1,000 $D_s\rightarrow \tau$ decays in $2.3 \times 10^8$ proton interactions in tungsten target to study the differential production cross section of $D_s$ mesons. State-of-the-art emulsion detectors with a nanometric-precision readout will be used to achieve this goal. This provides a unique opportunity to detect the peculiar double-kink topology of $D_s\rightarrow \tau\rightarrow X$ decays. 
The necessary detector technology, particularly the high-speed and high-precision readout of emulsions, has improved recently and become readily available. 
With this project, the uncertainty of the $\nu_\tau$ flux prediction will be reduced to under 10\%. Then, the systematic uncertainty of the $\nu_\tau$ cross-section measurement will be sufficiently low to test lepton universality in neutrino scattering.

Once $\nu_\tau$ production is established, the next stage will be to increase the data from $\nu_\tau$ detected events. We believe that both emulsion and liquid-argon detectors could successfully perform the measurements. This could be done within the framework of the SHiP project \cite{ship} at CERN because its beamline (beam-dump type) is well suited to this task. The outcome of our project will provide essential input into future $\nu_\tau$ experiments where high $\nu_\tau$ statistics are expected.

The measurement of the $\nu_\tau$ cross section is not only of interest for the fundamental study of lepton universality, it has also a practical impact on neutrino oscillation experiments. Mass hierarchy measurements by means of atmospheric neutrinos (for example, by Super-K and Hyper-K) rely on $\nu_e$ measurements that are contaminated by $\nu_\tau$ due to $\tau\rightarrow e$ decays. The systematic uncertainty from the $\nu_\tau$ cross section will be the limiting factor in their analysis \cite{wendell}. Accelerator-based experiments (DUNE \cite{dune}, Hyper-K \cite{hyper-k}) also suffer from $\nu_\tau$ background to $\nu_e$. Unlike other error sources, the uncertainty of the $\nu_\tau$ cross section cannot be constrained by near detector measurements, and this is relevant for CP-violation and mass hierarchy analyses. Along with the low-energy neutrino study ($E_\nu\simeq 1$ GeV), DUNE plans to extend its physics program to high-energy neutrino beams and study the $\nu_\tau$ appearance channel ($E_\nu> 10$ GeV). In this scenario, a better understanding of the $\nu_\tau$ cross section will be necessary. Results from this project will also benefit high-energy astrophysical $\nu_\tau$ observations (for example, IceCube \cite{icecube}).

\subsection{By-products}

With this project, we will analyse 2.3 $\times$ 10$^8$ proton interactions with an automated procedure. This, combined with the expected high yield of $10^5$ charmed decay products, will allow us to extract many more physical quantities. For example, we will measure the interaction lengths of charmed hadrons; this has never been accomplished and the results could be compared with the theoretical prediction. Another example is the detection of super-nuclei, which is an atomic nuclei involving charm quarks. Three candidate events of super-nuclei have been reported in the past using a similar method \cite{supernuclei}. However, no conclusive observation has been established. With 2 orders of magnitude more statistics, we will be in the position to make an important discovery. We would also measure $\Lambda_c$ production rates and topological branching ratios, improving current knowledge. In addition, the detector technology being developed for this project would be useful for future open charm measurements in heavy ion collisions.

\section{Principle of the experiment}

\begin{figure}[htbp]
\begin{center}
\includegraphics[width=\textwidth]{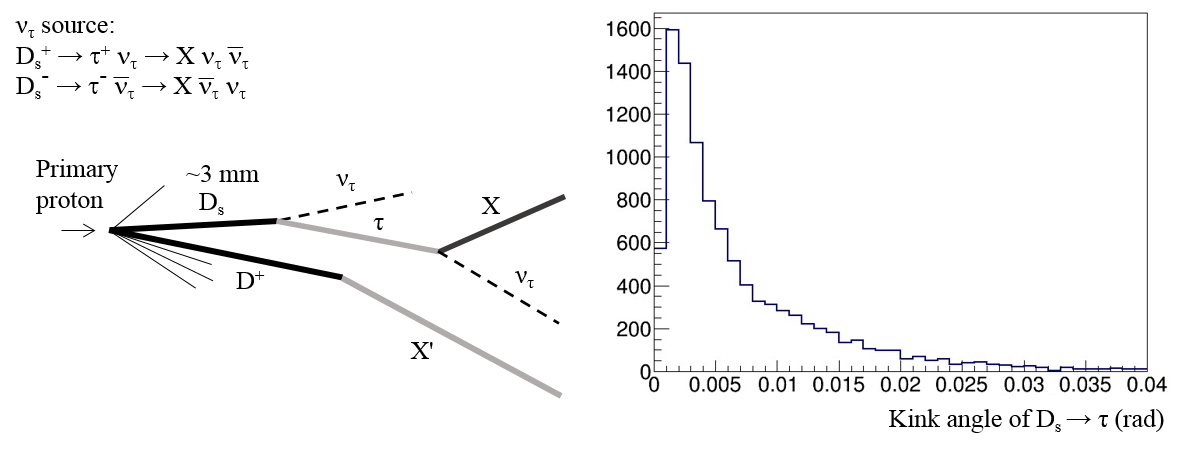}
\caption{Topology of $D_s\rightarrow\tau\rightarrow X$ events (left) and simulated kink angle distribution of $D_s\rightarrow\tau$ (right).}
\label{topology}
\end{center}
\end{figure}

The dominant source of $\nu_\tau$ is a sequential decay of $D_s$ mesons, $D_s^+ \rightarrow \tau^+ \nu_\tau \rightarrow X \nu_\tau \overline{\nu}_\tau$ and $D_s^- \rightarrow \tau^- \overline{\nu}_\tau \rightarrow X \overline{\nu}_\tau \nu_\tau$. These are created in high-energy proton interactions. The topology of such an event is shown in Figure \ref{topology} (left). $D_s$ decays to $\tau$ with a mean flight length of 3.3 mm and $\tau$ decays with a mean flight length of 2.0 mm. In addition, because charm quarks are created as a pair, another decay of charged/neutral charmed particles will be observed with a flight length of a few millimetres. This double-kink plus decay topology at this scale is a very peculiar signature for this process, and the background for such a topology is marginal. Measuring this signature has an advantage over comparing the $D_s$ production cross section and the decay-branching ratio of $D_s \rightarrow \tau$ separately. Our measurement gives an inclusive measurement of both of them; therefore some of the systematic errors are cancelled out. We therefore consider this measurement a direct measurement of $\nu_\tau$.

This measurement has a technical challenge. First, all the decays take place at a scale of millimetres. Second, the kink angle of $D_s \rightarrow \tau$ is anticipated to be very small, as shown in Figure \ref{topology} (right). The mean kink angle of $D_s$ is only 7 mrad. Detecting the small kink in a small volume is challenging. To detect this signal, we use an emulsion particle detector that provides the best spacial resolution among all particle detectors. Emulsion detectors and the automated readout systems have been successfully employed by several neutrino experiments such as CHORUS \cite{chorus}, DONUT  \cite{donut0} \cite{donut} and OPERA \cite{opera}.

\begin{figure}[htbp]
\begin{center}
\includegraphics[width=\textwidth]{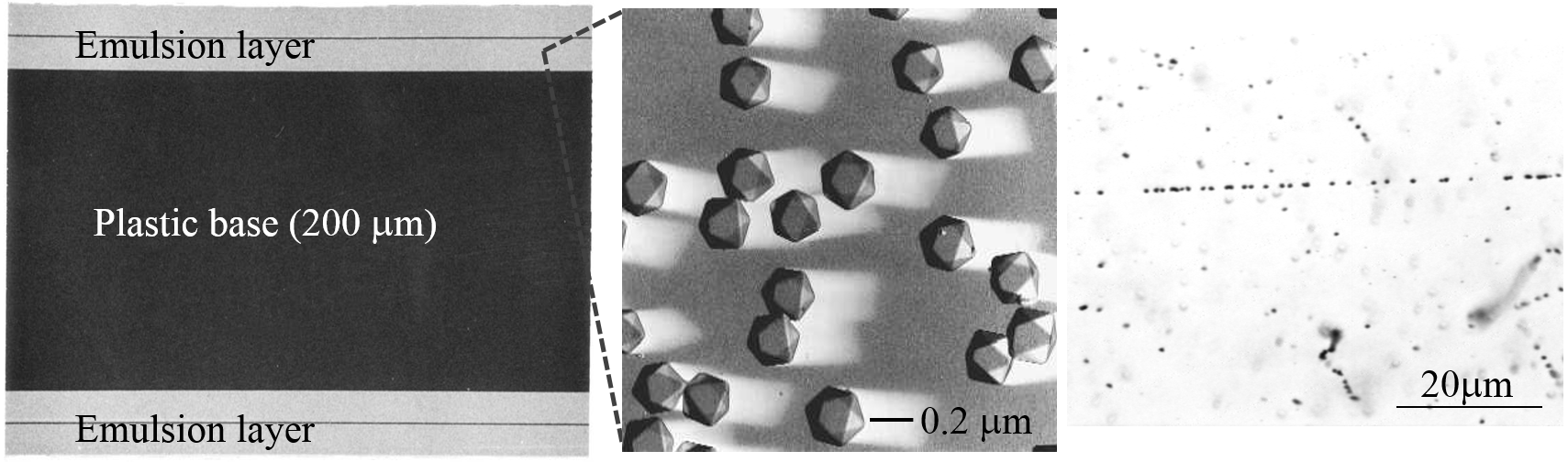}
\caption{Cross-sectional view of an emulsion film \cite{opera_film} (left), electron microscope picture of silver halide crystals (middle) and a minimum ionising track from a 10 GeV/c $\pi$ beam (right).}
\label{emulsion_film}
\end{center}
\end{figure}

Figure \ref{emulsion_film} (left) shows the cross-sectional view of a standard emulsion film \cite{opera_film} used in the OPERA experiment. The film comprises two emulsion layers (44 $\mu$m thick) that were poured onto both sides of a 200 $\mu$m thick plastic base; the film has a surface of approximately 12.5 cm $\times$ 10 cm. The emulsion comprises a number of silver bromide crystals, which are semiconductors (with a band gap of 2.5 eV) dispersed in gelatine. The diameter of these crystals is 0.2 $\mu$m, as shown in Figure \ref{emulsion_film} (middle). 
Once a charged particle passes through, the ionisation is recorded quasi-permanently, and then amplified and fixed by the chemical process. A minimum ionising track is shown in Figure \ref{emulsion_film} (right).
An emulsion detector with 200-nm crystals has a position resolution of 50 nm \cite{intrinsic_resolution}. An emulsion film structure with a 50-\si{\micro m}-thick sensitive layer on each side of a transparent 200-\si{\micro m}-thick plastic base has an intrinsic angular resolution of 0.35 mrad \footnote{
The hits near the plastic base are free from the distortion effect of the emulsion layer. Using two hits on the top and the bottom of the base gives a high resolution angle measurement, $\sigma_{\theta}^\textnormal{proj} = \sqrt{2}\cdot 50\textnormal{nm}/200\si{\micro m}$ = 0.35 mrad.
}.
This will allow us to set a kink detection threshold of 2 mrad at the 4 $\sigma$ confidence level.

\section{Detector structure}

To detect 1,000 $D_s \rightarrow \tau \rightarrow X$ events, $2.3 \times 10^8$ proton interactions need to be analysed; this is achieved by considering the detection efficiency to be 20\% (see section 5), the branching fraction of $D_s\rightarrow\tau$ (5.48 $\pm$ 0.23)\% \cite{pdg2017} and the $D_s$ production cross section in the tungsten target approximately 4 $\times$ 10$^{-4}$ per proton interaction 
\footnote{
$\sigma$($D_s$) is not directly measured for the 400-GeV proton beam. We get the ratio $\sigma$($D_s$)/$\sigma$($D^0$) = 0.21 for 230-350 GeV pion or proton beams \cite{NA32,E769_1,WA92} and assume that it does not depend on the beam energy. $\sigma$($D^0$) is measured to be 18.3 $\mu$b/nucl for a 400-GeV proton beam \cite{NA27}. Thus, the $D_s$ production rate per proton interaction is estimated to be (184$^{0.99}$ $\times$ 3.78 $\mu$b)/(184$^{0.71}$ $\times$ 40 mb) = 4$\times$10$^{-4}$.
}.

\begin{figure}[htbp]
\begin{center}
 \includegraphics[width=\textwidth]{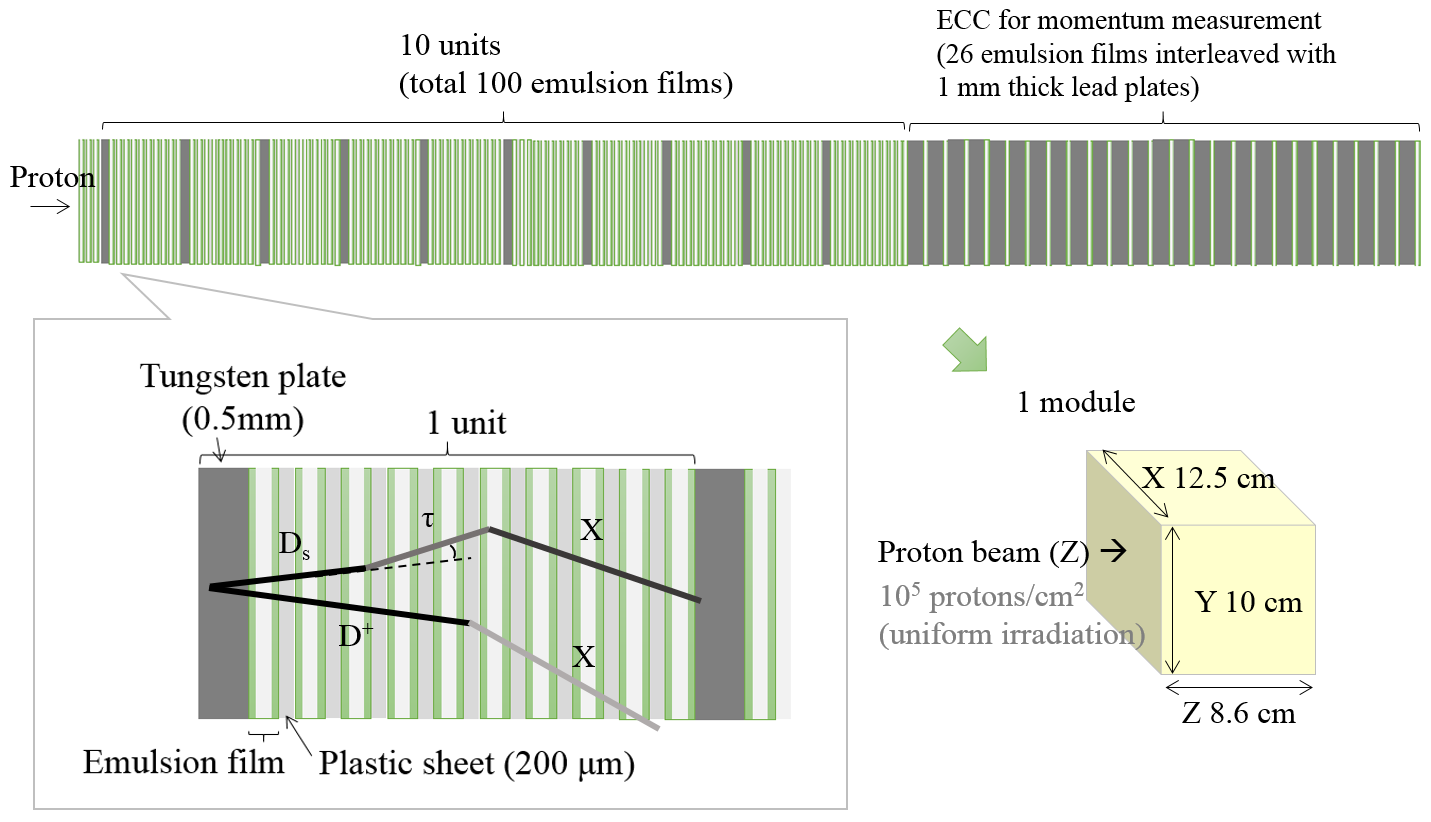}
\caption{Schematic of the module structure. A tungsten plate, the proton interaction target, is followed by 10 emulsion films and 9 plastic sheets as a tracker and decay volume. The sensitive layers of emulsion detectors are indicated in green. This structure is repeated 10 times; then, a so-called Emulsion Cloud Chamber (ECC) structure follows for momentum measurement of the daughter particles.}
\label{module}
\end{center}
\end{figure}

The structure of the detector unit is shown in Figure \ref{module}. A 500-\si{\micro m}-thick tungsten target is followed by 10 emulsion films interleaved with 200-\si{\micro m}-thick plastic sheets which act as a decay volume for short-lived particles as well as high-precision particle trackers. This unit structure is repeated 10 times to construct a module. A so-called Emulsion Cloud Chamber (ECC), which has repeated structure of emulsion films interleaved with 1-mm-thick lead plates, follows for momentum measurement of the daughter particles. Momenta of the reconstructed tracks will be determined by Multiple Coulomb Scattering (MCS) in the ECC \cite{mcs}. Three additional emulsion films will be placed upstream to tag the incoming protons. A single module is 12.5 cm wide, 10 cm high and 8.6 cm thick (for a total of 129 emulsion films). With this module, $4.6 \times 10^9$ protons on target are needed to accumulate $2.3 \times 10^8$ proton interactions in the tungsten plates.

\begin{figure}[htbp]
\begin{center}
\begin{tabular}{cc}
\includegraphics[height=5.5cm]{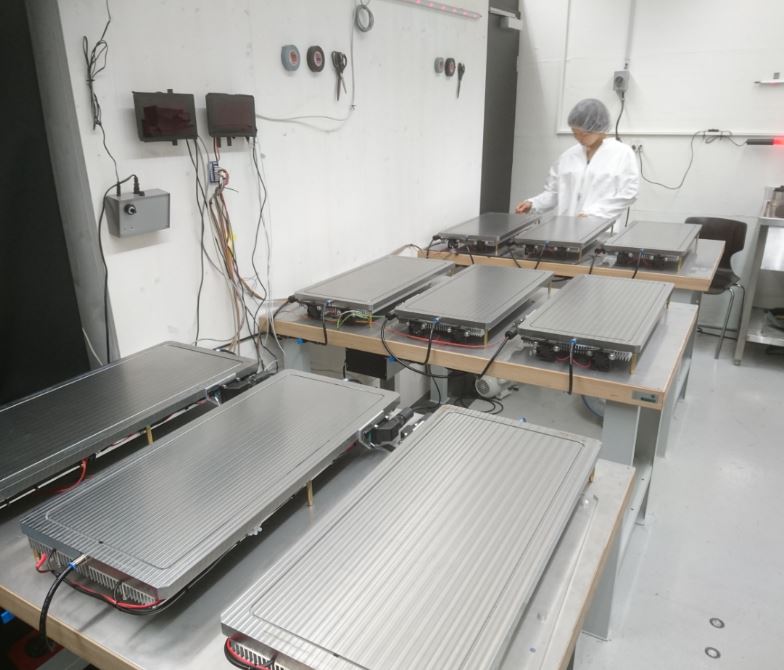}&
\includegraphics[height=5.5cm]{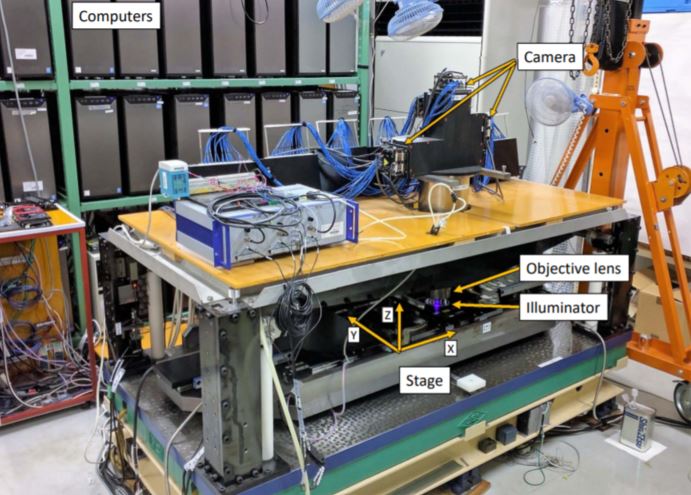}
\end{tabular}
\caption{The emulsion detector production facility at 70 m water-equivalent underground (left) and the fast emulsion readout system, Hyper Track Selector \cite{hts2} (right).
}
\label{facilities}
\end{center}
\end{figure}

The emulsion detector and its readout system can operate at a track density up to about $10^6$ particles/\si{cm^2}, which limits the proton density at the upstream surface of the module to less than $10^5$/\si{cm^2}. Therefore, it is envisaged that 368 modules will collect $2.3 \times 10^8$ interactions. The total module surface is 4.6 \si{m^2}, which corresponds to an emulsion film area of 593 \si{m^2}. The emulsion detector production facilities have been established at Nagoya University and the University of Bern. Both facilities can produce emulsion films with a rate of 10 \si{m^2} per week. The 70 m water-equivalent underground laboratory in Bern is shown in Figure \ref{facilities} (left); it would be used for the production and storage of the emulsion films, together with the facility in Nagoya. The production of 593 \si{m^2} emulsion film would take about eight months. We are pursuing industrial mass production by companies, which would condiderably shorten the production time.

\section{Readout and data analysis of the emulsion modules}

Owing to the requirements of detecting small kinks in $D_s\rightarrow\tau$ decays, the analysis will be performed in stages: (1) full area scanning by a fast system with relatively poor angular resolution, (2) detection of $\tau\rightarrow X$ decays ($\tau$ decay trigger, $\theta_{\tau\rightarrow X}^\textnormal{kink}\simeq$100 mrad) and (3) precision measurement around the $\tau$ decay to find $D_s\rightarrow\tau$ decays ($\theta_{D_s\rightarrow\tau}^\textnormal{kink}\simeq 7$ mrad) by a high-precision scanning system. 

Fast scanning will be performed by the Hyper Track Selector (HTS) system \cite{hts1,hts2} of the Nagoya University group. The system is shown in Figure \ref{facilities} (right). 
It has a scanning speed of 0.5 \si{m^2/h/layer}. The total emulsion area of 1186 \si{m^2} (593 \si{m^2} films $\times$ two sensitive layers) can be scanned in six months. 

\begin{figure}[htbp]
\begin{center}
\begin{tabular}{cc}
\hspace{-5mm}\includegraphics[height=5.5cm]{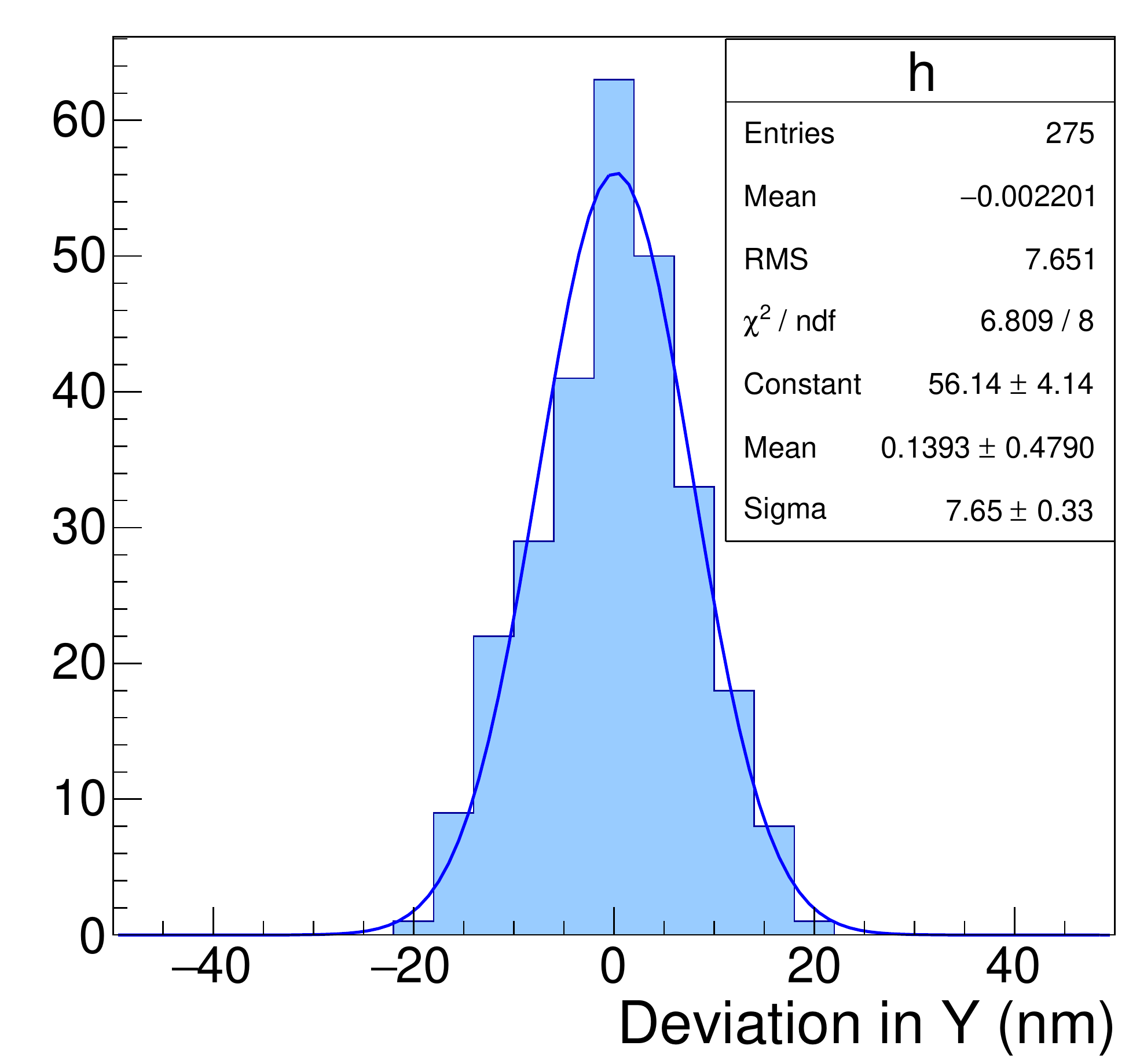}&
\hspace{-4mm}\includegraphics[height=5.5cm]{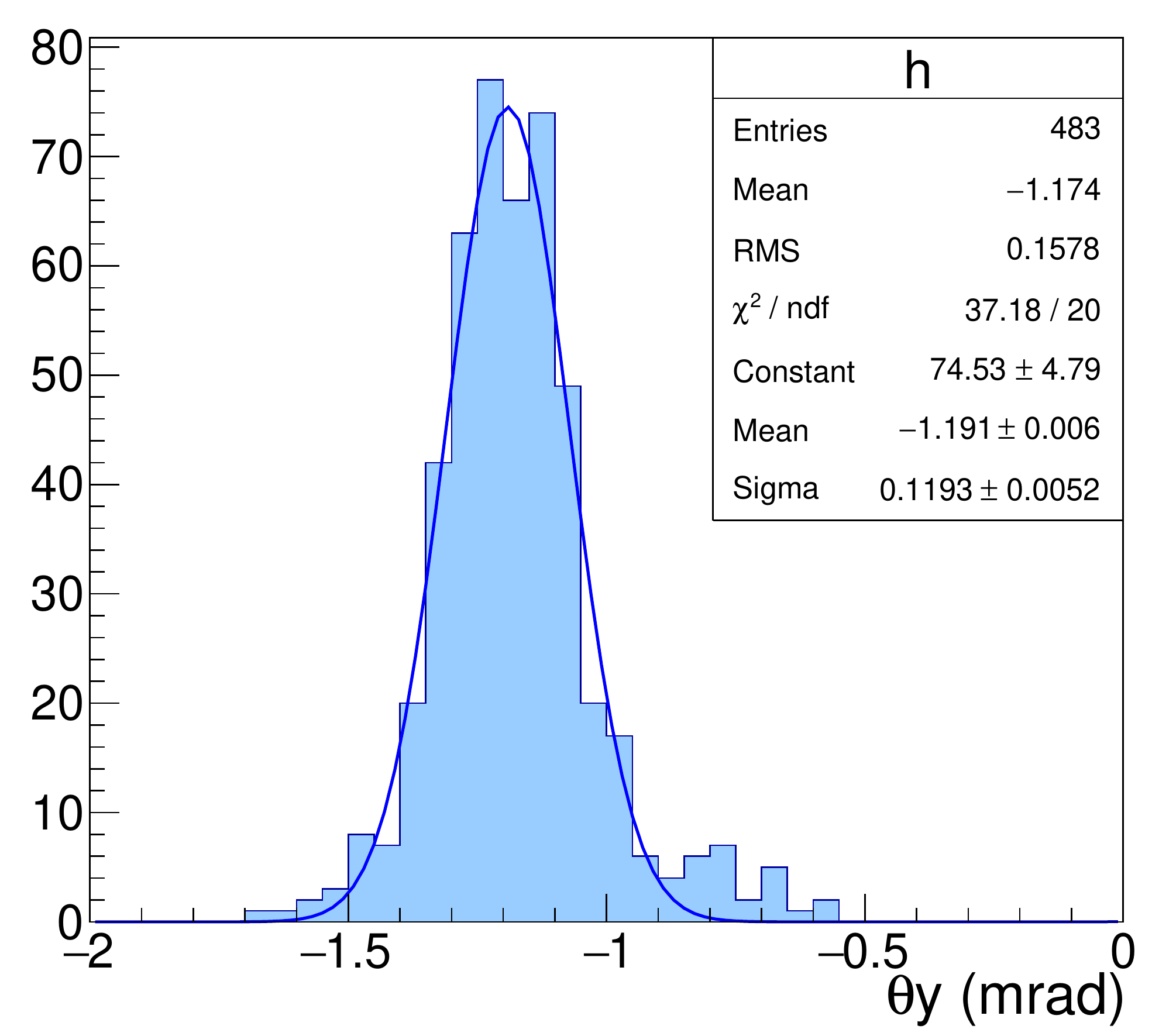}\\
\end{tabular}
\hspace{7mm}\caption{Readout resolutions by the high-precision system in Bern. The same hit (left) or track (right) was measured multiple times and reproducibilities were evaluated.
}
\label{resolution}
\end{center}
\end{figure}

After the $\tau$ decay trigger, the events will be analysed in Bern \cite{atmic} (and possibly other sites with a high precision system) with a piezo-based high-precision Z axis, allowing measurement of hits in emulsion with a nanometric resolution. As shown in Figure \ref{resolution}, the position measurement reproducibility
for a single hit (grain) is below 10 nm, which is better than the 50-nm intrinsic resolution of the emulsion detector. The angular measurement reproducibility is found to be 0.15 mrad (RMS), including measurements' effects (such as skewing, hysteresis, encoding resolution of X and Y axes and optical aberration). This readout resolution would be sufficient to detect $D_s\rightarrow\tau$ kinks with high reliability.

Issues related to the planarity of emulsion films (angular alignment between films) and the distortion of emulsion layer could be additional sources of error. However, there are high-energy proton tracks with high density (100 tracks in every microscope view of 300 $\mu$m $\times$ 400 $\mu$m) and each of them would give 0.35 mrad angular correction. The planarity and distortion would effectively be cancelled by performing a relative angular measurement between the proton tracks.  

\section{Expected performance}

The $\tau$ lepton decays to a single charged particle (1-prong) with a branching ratio of 85\% and 3-prong with a branching ratio of 15\% \cite{pdg2017}. The following estimation was performed for the case of $\tau\rightarrow$1-prong. 

The detection efficiency is estimated to be 20\% using a PYTHIA 8.1 \cite{pythia81} simulation, which required the following preliminary criteria to be fulfilled: 
(1) the parent has to pass through at least one emulsion film (two sensitive layers), 
(2) the first kink daughter has to pass through at least two sensitive layers and the kink angle is $\geq$ 2 mrad, 
(3) the flight length of the parent and the first kink daughter has to be $<$ 5 mm, 
(4) the second kink angle is $\geq$ 15 mrad and 
(5) the partner of the charm pair is detected with 0.1 mm $\leq$ flight length $<$ 5 mm (they can be charged decays with a kink angle >15 mrad or neutral decays).

\begin{figure}[htbp]
\begin{center}
\begin{tabular}{cc}
\includegraphics[height=7cm]{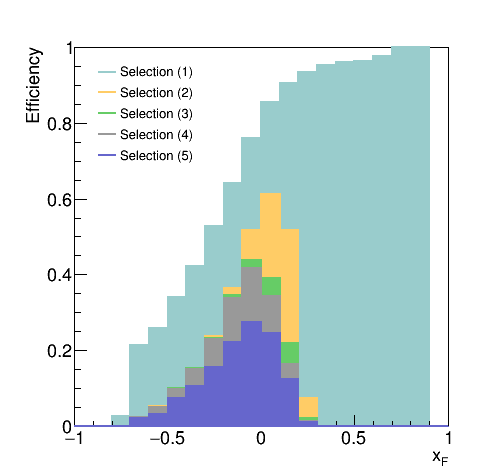}&
\includegraphics[height=7cm]{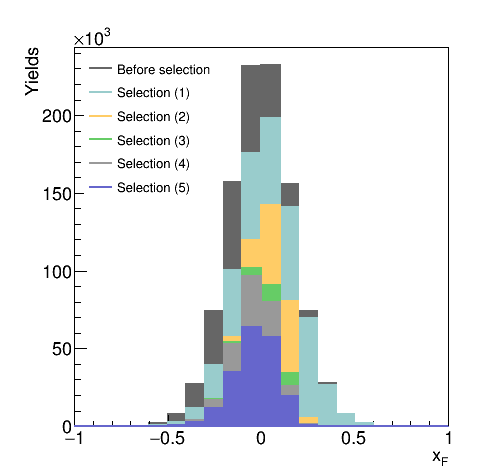}
\end{tabular}
\caption{The estimated detection efficiency as a function of $x_F$ (left) and the $x_F$ distributions after the selection (right). Selection (1)-(5) are described in the text.
}
\label{fig:eff}
\end{center}
\end{figure}

\begin{figure}[htbp]
\begin{center}
\begin{tabular}{cc}
\includegraphics[height=5cm]{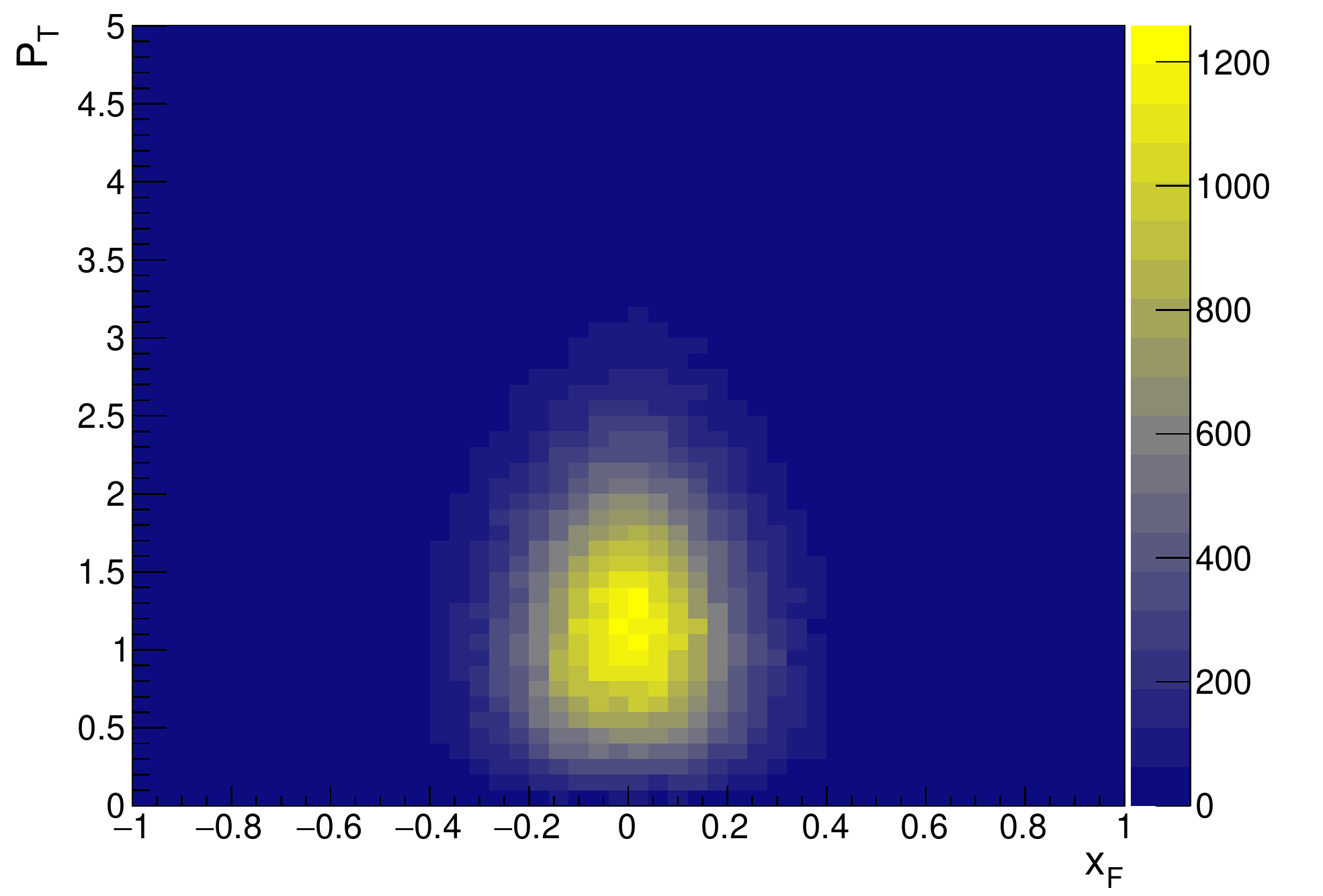}&
\includegraphics[height=5cm]{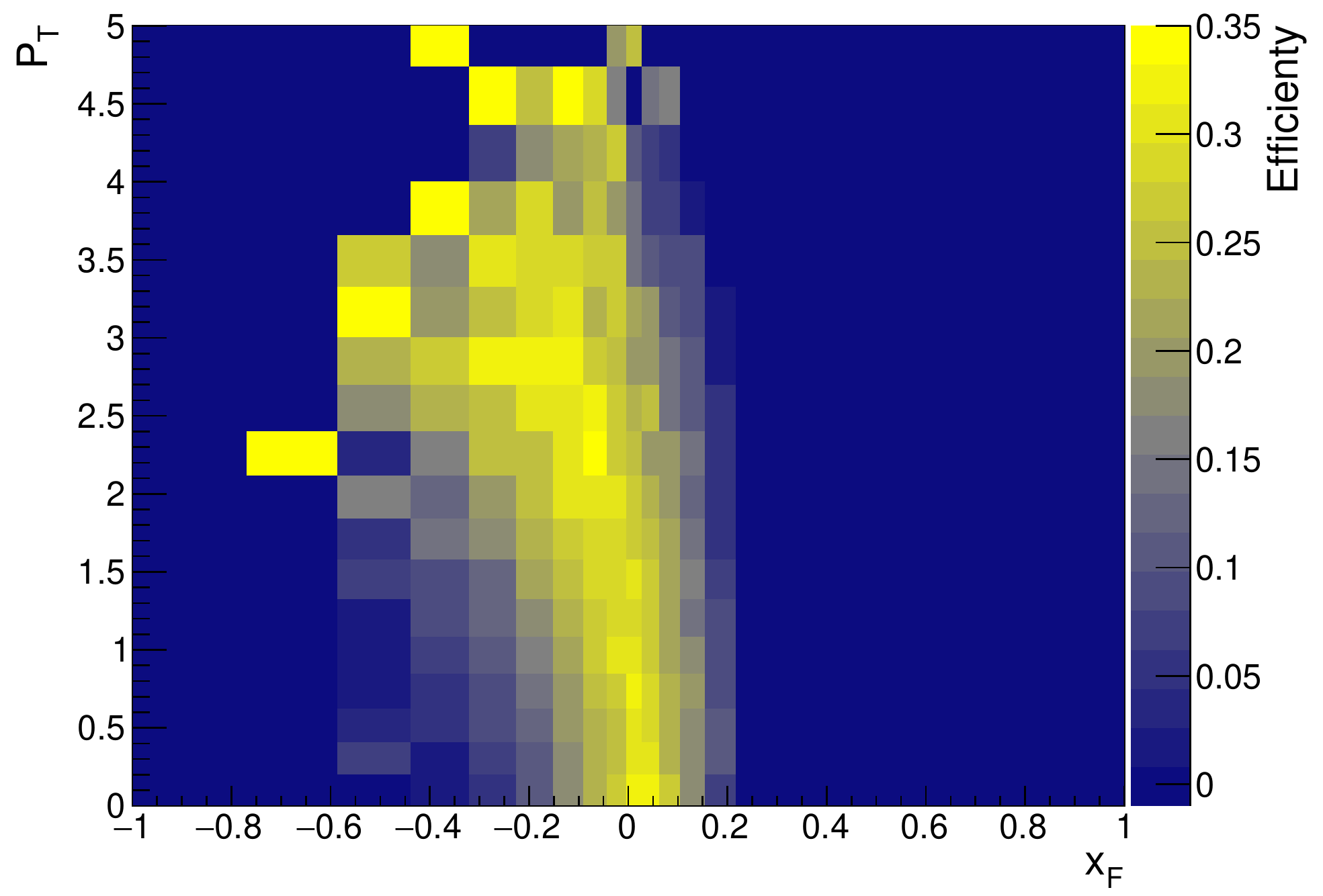}
\end{tabular}
\caption{The $x_F$-$p_T$ distribution before the selection (left) and the estimated detection efficiency in the $x_F$-$p_T$ plane (right).
}
\label{fig:xf-pt}
\end{center}
\end{figure}

Table \ref{tb:eff} shows the breakdown of the efficiency estimation. Figure \ref{fig:eff} shows the estimated detection efficiency as a function of the Feynman $x$ ($x_F=2p^{CM}_Z/\sqrt{s}$) (left) and the $x_F$ distributions after the selection (right). Figure \ref{fig:xf-pt} shows the $x_F$ - $p_T$ distribution before the selection (left) and the detection efficiency in the $x_F$ - $p_T$ plane (right). Further improvements of the detection efficiency will be studied. Because the angular resolution is affected by the length that the particles pass through, better resolutions can be achieved when the particles pass through more than two sensitive layers. The efficiency could be improved by applying different thresholds for the first kink angle, depending on the flight length of the parent and daughter.

\begin{table}[hbtb]
\begin{center}
\begin{tabular}{lr}
\hline
Selection & Total efficiency (\%)\\
\hline
(1) Flight length of $D_s$ $\geq$ 2 emulsion layers &  77\\
\hline
(2) Flight length of $\tau$ $\geq$ 2 emulsion layers and $\Delta\theta_{D_s\rightarrow\tau}\geq$ 2 mrad & 43 \\
\hline
(3) Flight length of $D_s$ $<$ 5 mm and flight length of $\tau$ $<$ 5 mm & 31 \\
\hline
(4) $\Delta\theta_{\tau \rightarrow X}\geq$ 15 mrad & 28 \\
\hline
(5) Pair charm: 0.1 mm $\leq$ flight length $<$ 5 mm & 20 \\
(charged decays with $\Delta\theta\geq$ 15 mrad or neutral decays) & \\
\hline
\end{tabular}
\caption{
Breakdown of the efficiency estimation.}
\label{tb:eff}
\end{center}
\end{table}

The main background source comprises hadronic secondary interactions. These can be reduced by requesting the absence of nuclear fragments (either in the backward or forward hemisphere). We performed a preliminary estimate as follows: the mean free path in an emulsion for interactions without any detectable nuclear fragments is 11 m for a 5-GeV pion beam \cite{charon} (this value has yet to be measured for plastics). Using this value as an estimate for both emulsions and plastic sheets, the probability of having a double kink during a flight length $<$ 5 mm is $(4.5 \times 10^{-4})^2$ per particle. The particle multiplicity is expected to be approximately 10; therefore, the probability of having a double kink that mimics a $D_s\rightarrow\tau$ event as well as a kink that mimics a partner charm event is $10 \times (4.5 \times 10^{-4})^2 \times 10 \times (4.5 \times 10^{-4}) = 9 \times 10^{-9}$. This is acceptable because it is smaller than the signal probability of $5 \times 10^{-6}$ per proton interaction. A more careful simulation based on FLUKA \cite{fluka} is being performed to precisely estimate the overall background and study any possible improvements through multivariate techniques. 
The estimation of hadronic background can be validated through real data obtained by analysing a region outside of the decay volume where no signal event is expected.

\begin{figure}[htbp]
\begin{center}
\begin{tabular}{c}
\includegraphics[height=6.0cm]{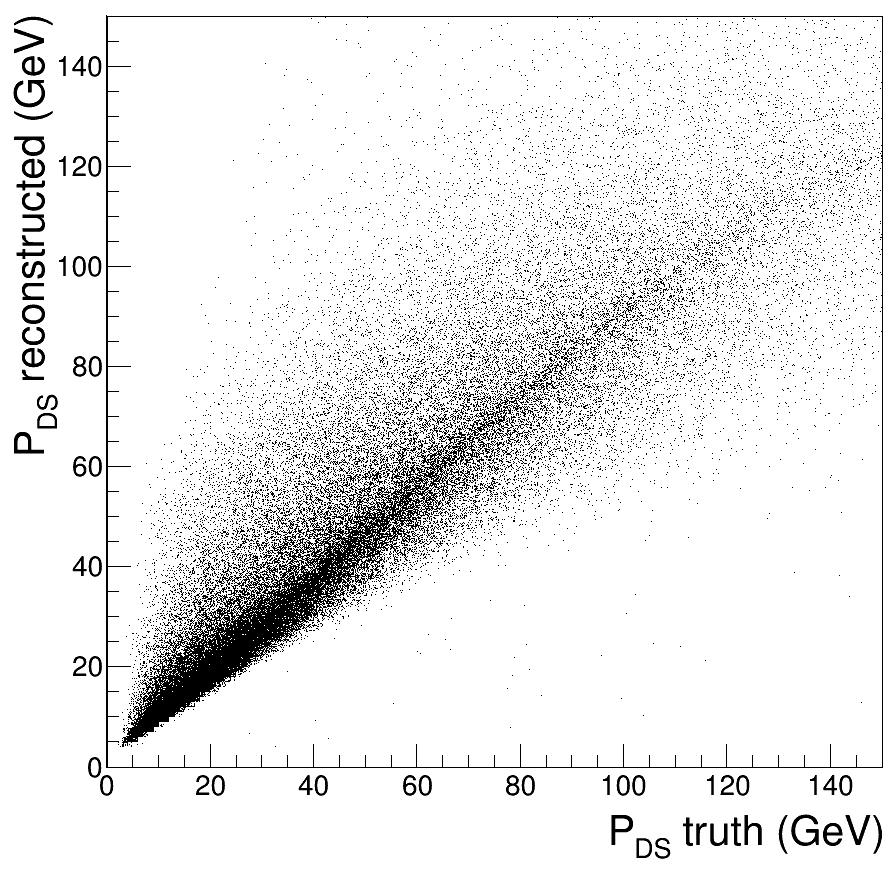}
\end{tabular}
\caption{Reconstructed $D_s$ momenta versus true momenta. The momenta are reconstructed by a machine-learning algorithm using the topological variables $\theta_{D_s\rightarrow\tau}$, $\theta_{\tau\rightarrow X}$, $FL_{D_s\rightarrow\tau}$, $FL_{\tau\rightarrow X}$. A gaussian fit of $\Delta P/P$ provides a $\sigma$ of 18\%.}
\label{dsmom}
\end{center}
\end{figure}

To study the differential production cross section of $D_s$, the momentum of $D_s$ ($P_{D_s}$) must be known. Because the $D_s$ decays quickly and the invisible $\nu_\tau$ escapes measurement, the direct measurement of $P_{D_s}$ is not possible. However, the peculiar event topology gives us indications of $P_{D_s}$. For example,  $D_s \rightarrow \tau \nu_\tau$ is a two-body decay having a well-defined decay momentum. Therefore, the kink angle of $D_s \rightarrow \tau$ is a good approximation of $P_{D_s}$. Because the $D_s \rightarrow \tau \rightarrow X$ decay topology has two kink angles ($\theta_{D_s\rightarrow\tau}$, $\theta_{\tau\rightarrow X}$) and two flight lengths ($FL_{D_s\rightarrow\tau}$, $FL_{\tau\rightarrow X}$), a combination of these variables effectively provides an estimate of $P_{D_s}$. A machine-learning algorithm was trained with a simulated sample ($\tau \rightarrow$ 1 prong) using the four variables to obtain $P_{D_s}$. The result is shown in Figure \ref{dsmom}. The momentum resolution was estimated to be 18\%.


\begin{figure}[htbp]
\begin{center}
\begin{tabular}{c}
\includegraphics[height=6cm]{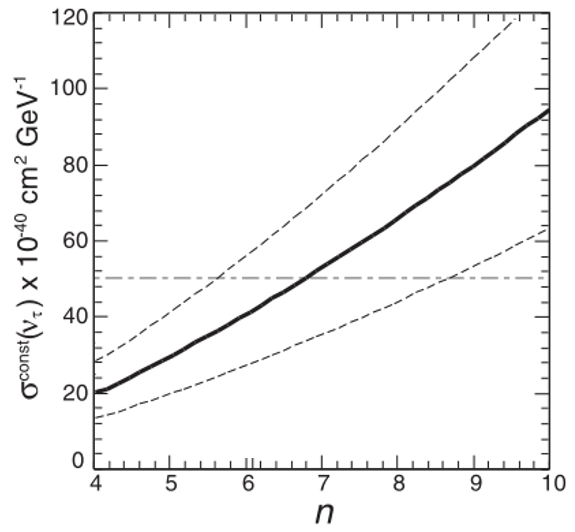}
\end{tabular}
\caption{The energy-independent $\nu_\tau$ cross section as a function of the parameter $n$, measured in DONUT \cite{donut}, is represented by the thick solid curve. The 1-$\sigma$ statistical error is shown by the dashed curves and the average of the standard model neutrino and antineutrino cross sections is shown by the dot-dashed horizontal line.
}
\label{fig:donut_fig16}
\end{center}
\end{figure}

The result of this project will improve the $\nu_\tau$ flux prediction. As described in Section 1, it is conventional to estimate the differential production cross section of $D_s$ by a phenomenological formula, d$^2$$\sigma$/(d$x_F$d$p_T^2$) $\propto$ (1-|$x_F$|)$^n$ $\exp$(-b$p_T^2$), using parameters $n$ and $b$. The DONUT result was a cross section as a function of the parameter $n$; the energy-independent part of the total CC cross section $\sigma_{\nu\tau}^{const}$ = 2.51 $n^{1.52}$ (1 $\pm$ 0.33 (stat.) $\pm$ 0.33 (syst.)) $\times$ 10$^{-40}$ cm$^2$GeV$^{-1}$. This parameter-dependent cross-section result is shown in Figure \ref{fig:donut_fig16}. To reduce the systematic uncertainty of the $\nu_\tau$ CC cross section lower than 10\%, the parameter $n$ of the $D_s$ production has to be determined at a precision of $\sim$ 0.4.

\begin{figure}[htbp]
\begin{center}
\begin{tabular}{c}
\includegraphics[height=6.2cm]{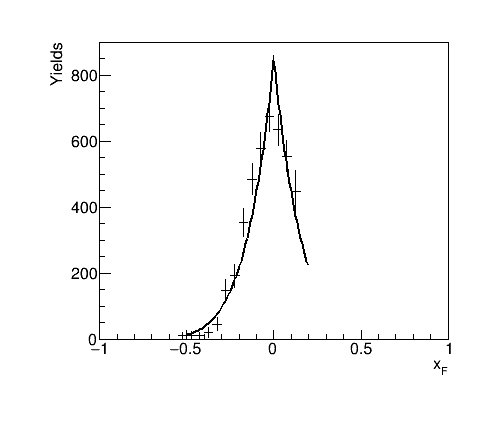}
\end{tabular}
\caption{Monte Carlo distribution of reconstructed x$_F$ for 1,000 events, corrected by the detection efficiency.
}
\label{xf_1000}
\end{center}
\end{figure}

\begin{figure}[htbp]
\begin{center}
\begin{tabular}{cc}
\includegraphics[height=6.2cm]{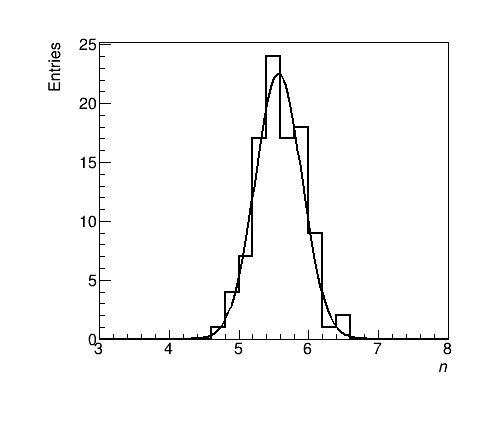}&
\raisebox{5mm}{\includegraphics[height=5.2cm]{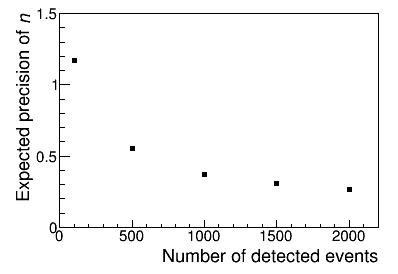}}
\end{tabular}
\caption{
Distribution of the parameter $n$ derived from the x$_F$ distributions of 1,000 events repeated for a hundred experiments (left). Expected precision for the measurement of parameter $n$ as a function of the number of detected events (right). 
}
\label{precision_n}
\end{center}
\end{figure}

Figure \ref{xf_1000} shows a Monte Carlo distribution of reconstructed x$_F$ for 1,000 events, corrected by the detection efficiency. The fit of the yields to $(1-|x_F|)^n$ gives a value for the parameter $n$. Repeating this experiment a hundred times, the distribution of the parameter $n$ is shown in Figure \ref{precision_n} (left). Figure \ref{precision_n} (right) shows the expected precision for the measurement of parameter $n$ as a function of the number of detected events. Precision of 0.4 could be achieved by detecting 1,000 events. 

\begin{figure}[htbp]
\begin{center}
\begin{tabular}{c}
\includegraphics[height=6cm]{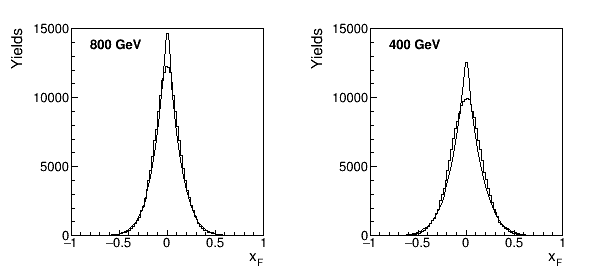}
\end{tabular}
\caption{$x_F$ distributions for $D_s^{\pm}$ production in 800 (left) and 400 GeV (right) proton-nucleon interactions. The fit of the yield to $(1-|x_F|)^n$ gives $n$ = 6.9 for 800 GeV and $n$ = 5.8 for 400 GeV.}
\label{xf_800_400}
\end{center}
\end{figure}

With the revised value of $n$, the $\nu_\tau$ CC cross section measured by DONUT is re-evaluated. To achieve the DONUT result, we extrapolate data from 400 to 800 GeV. Figure \ref{xf_800_400} shows $x_F$ distributions for $D_s^{\pm}$ production in 800 and 400 GeV proton-nucleon interactions predicted by PYTHIA 8.1 simulation. The fit of the yield to $(1-|x_F|)^n$ gives $n$ = 6.9 for 800 GeV and $n$ = 5.8 for 400 GeV. The uncertainty on scaling up from 400 to 800 GeV is less than the statistical uncertainty of the DONUT result. 
With future $\nu_\tau$ detection experiments using a 400-GeV proton beam, the total uncertainty of the $\nu_\tau$ CC cross-section measurement will be reduced.

\section{Beam tests and the exposure scheme}

Beam test experiments were performed in November 2016 at the H4 beamline and in May 2017 at the H2 beamline. The objective was to understand the exposure scheme, show the angular resolution and demonstrate decay topology detection. 

The emulsion films were produced at the production facility in Bern. The tungsten plates were obtained from the GoodFellow company. The thickness of the plates was measured to be 507 $\pm$ 2 \si{\micro m} with a dispersion of 16 \si{\micro m} among 67 plates. The assembly of the modules took place in the darkroom at CERN.

\begin{figure}[htbp]
\begin{center}
\begin{tabular}{cc}
\includegraphics[height=5.5cm]{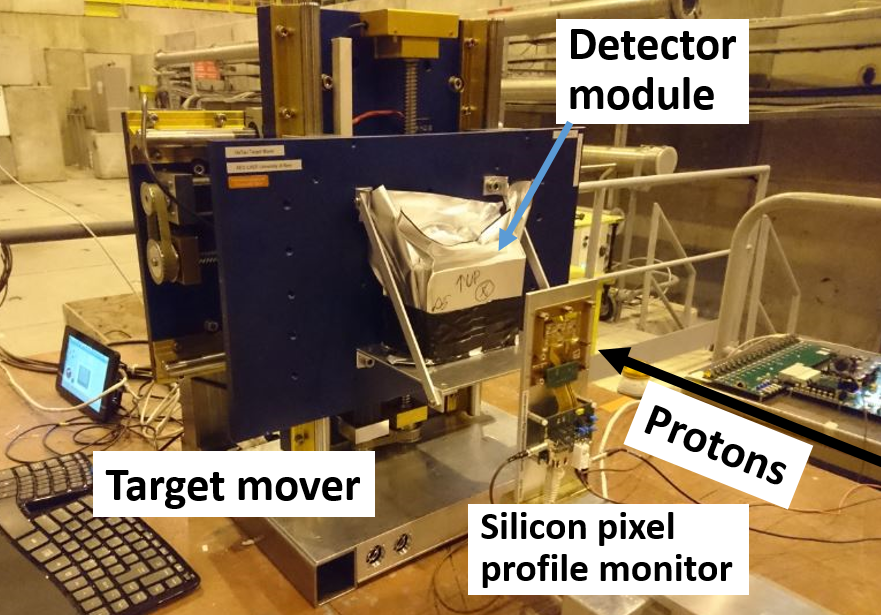}
\includegraphics[height=5.5cm]{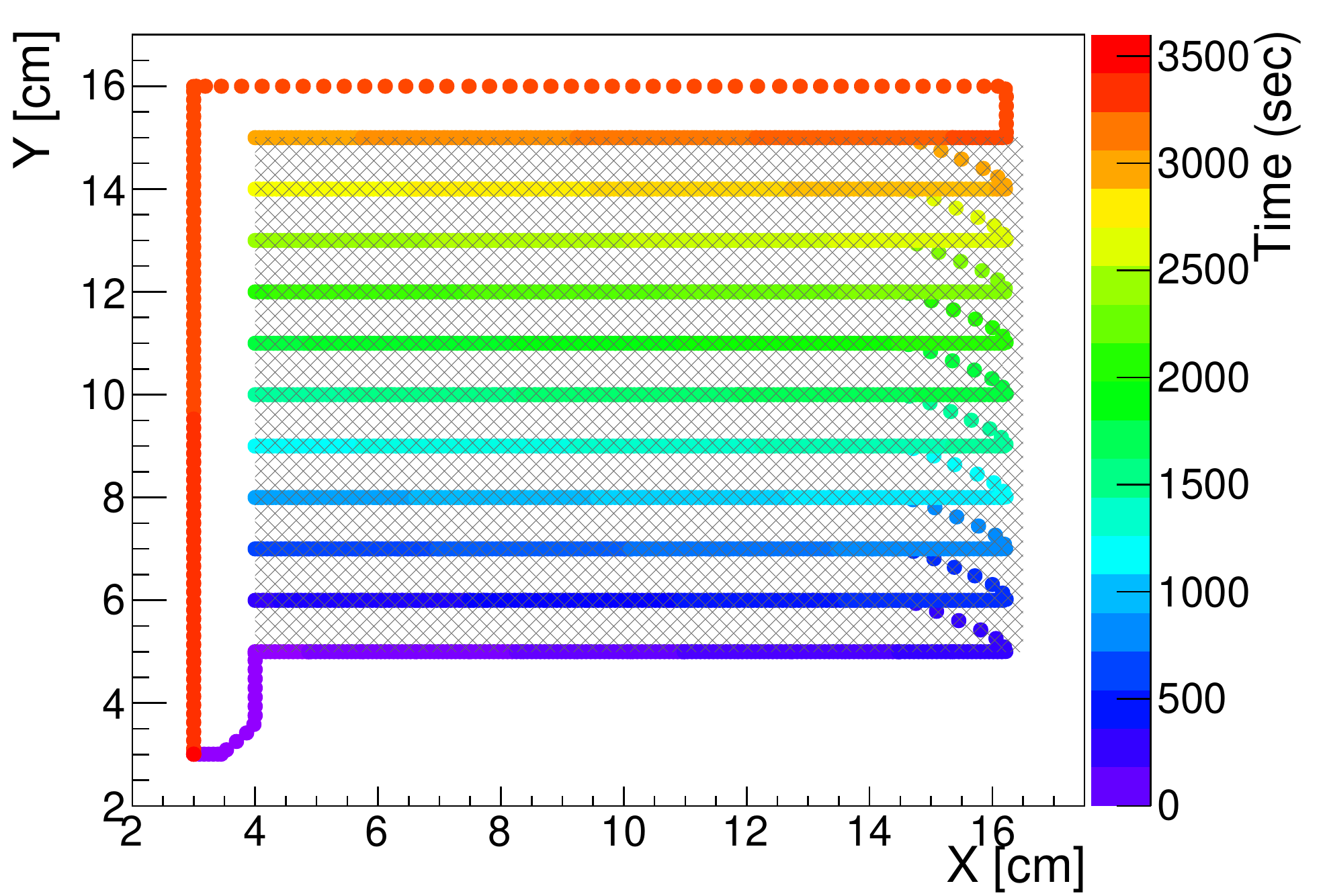}
\end{tabular}
\caption{Test beam set-up of the 2016 measurement campaign at CERN (left) and the scanning sequence of the target mover (right). The beam centre (coloured dots) with respect to the detector module surface (the hatched region) is shown. The colour axis shows time. 
}
\label{testbeam}
\end{center}
\end{figure}

The beam profile was monitored by the silicon pixel sensor telescope built in Bern comprising two planes of ATLAS IBL modules with FE-I4A front-end readout chips \cite{silicon_1,silicon_2}. The sensor dimension is 2.04 cm $\times$ 1.87 cm. It contains 80 $\times$ 336 pixels, each having a size of 250 $\mu$m $\times$ 50 $\mu$m. To realize a uniform irradiation on the detector module surface, the modules were mounted on a motorised X-Y stage (target mover, constructed in Bern) moving synchronously with the SPS spills. The experimental set-up at the beamline is shown in Figure \ref{testbeam}. 

\begin{figure}[htbp]
\begin{center}
\includegraphics[height=7cm]{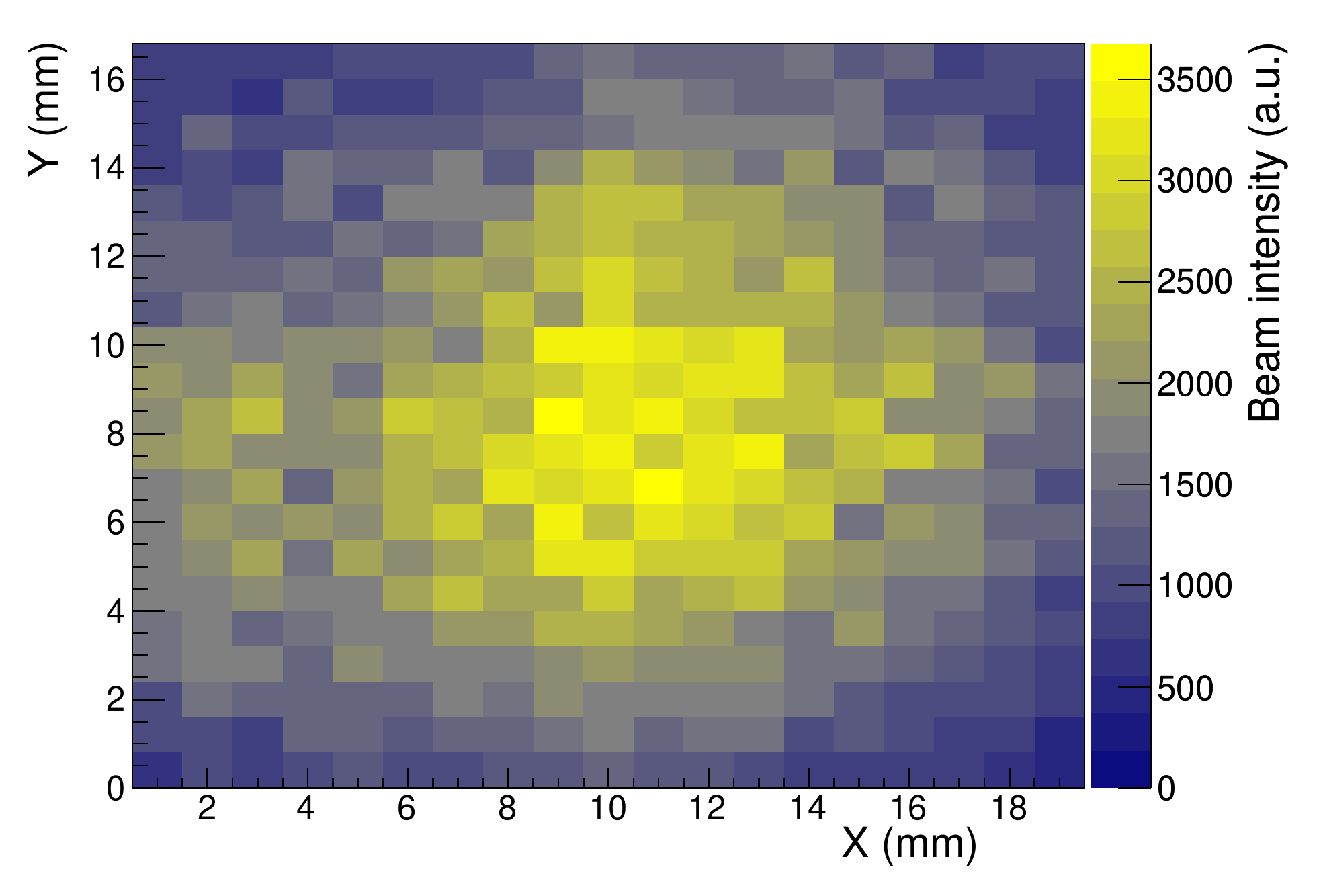}
\caption{Enlarged proton profile measured by the silicon pixel sensor. The parameters fitted by 2D Gaussian are $\sigma_{X}$ = 7.6 mm and $\sigma_{Y}$ = 6.0 mm.
}
\label{silicon_profile}
\end{center}
\end{figure}

The detector module was primarily scanned in the X direction with respect to the proton beam when the beam is on. Several spills are needed to scan one line, then the Y axis is moved by $Y_{step}$ and the next line is scanned as shown. 
To achieve a uniformity of proton density in the Y direction, $Y_{step}$ must satisfy a condition related to the proton-beam size ($\sigma_{y}$), $Y_{step}<1.7\sigma_y$. Typically, the beam size of protons, $\sigma_y$, at 400 GeV is only a few millimetres. Therefore, the beam size was enlarged to reduce the number of lines needing to be scanned. The proton-beam profile is shown in Figure \ref{silicon_profile}, which allows $Y_{step}$ = 1.0 cm to be set.

\begin{figure}[htbp]
\begin{center}
\includegraphics[width=\textwidth]{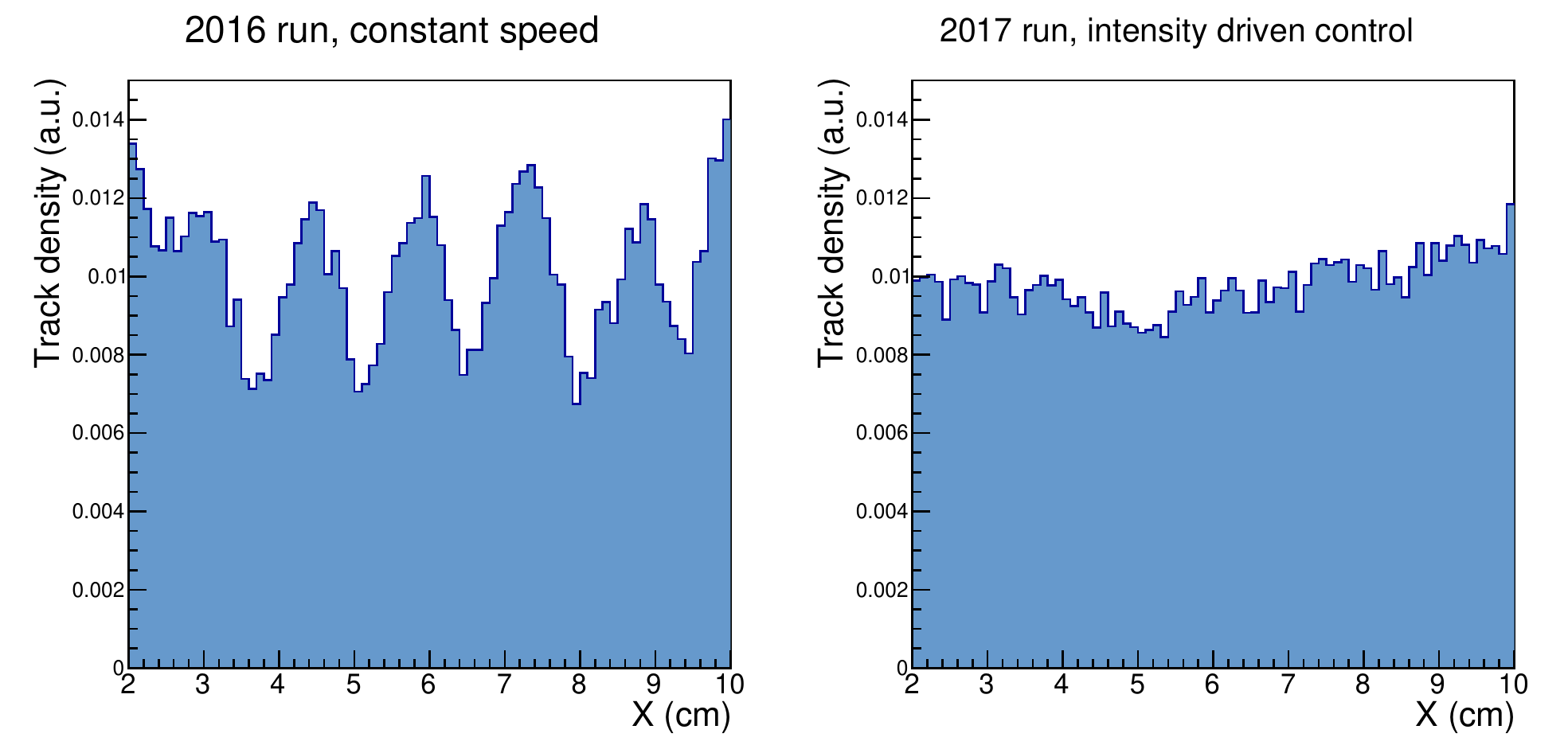}
\caption{The X position distribution of proton tracks in the detector; the detector module was moved with a constant speed. A wavy pattern was observed owing to the timing structure of the beam during FT (left) and an intensity-driven control was applied. The heterogeneity was at the acceptable level (right).}
\label{positiondist}
\end{center}
\end{figure}

\begin{figure}[htbp]
\begin{center}
\includegraphics[width=0.7\textwidth]{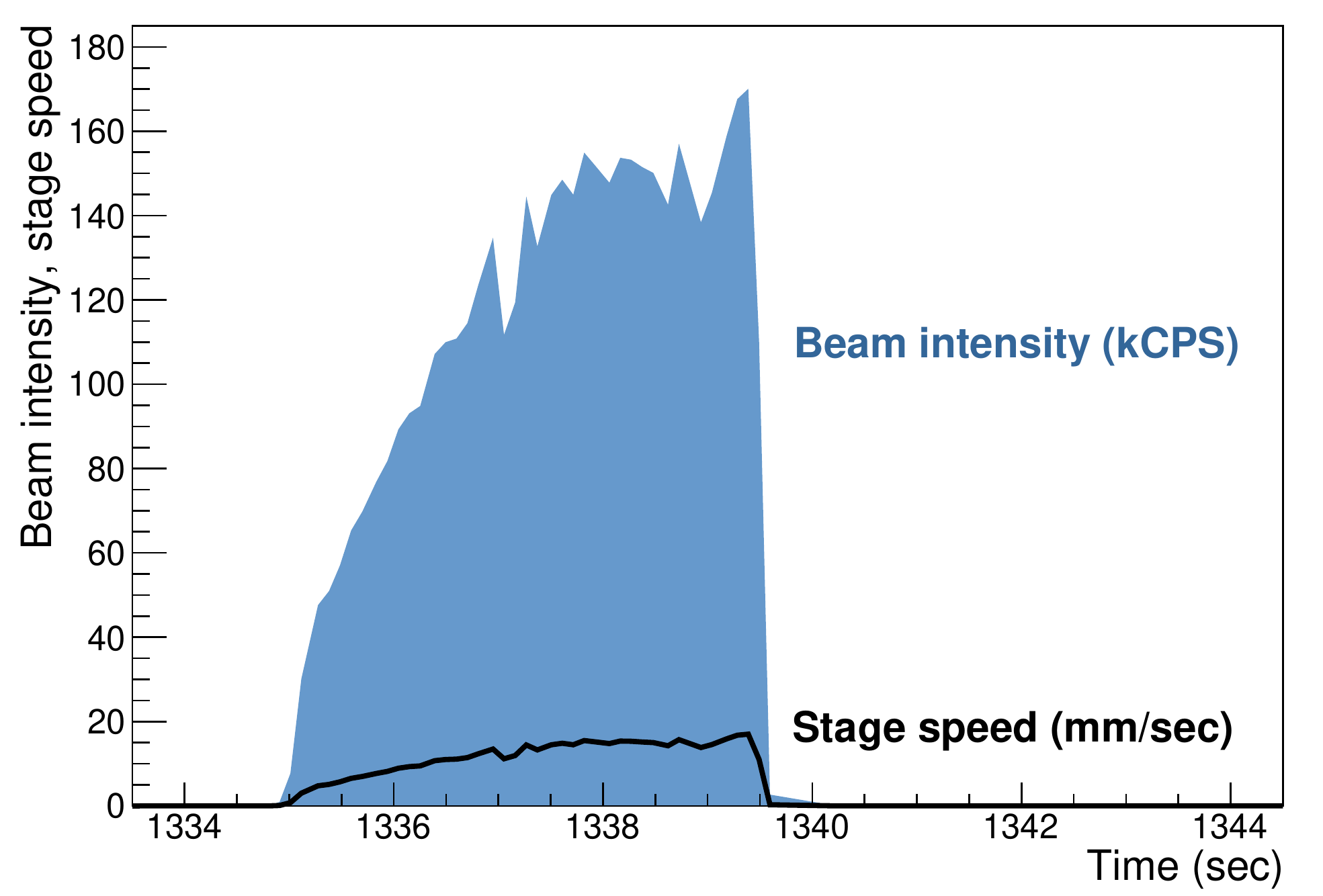}
\caption{The time profile of the proton beam in a spill measured by the scintillator counter at H2 during the 2017 run. The speed of the target mover was altered accordingly.}
\label{timeprofile}
\end{center}
\end{figure}

The distance to be moved for every spill ($X_{step}$) depends on the desired proton density ($\rho$ [\si{protons/cm^2}]) and the intensity of protons ($I_{protons}$[protons/spill])
\[
X_{step} = \frac{I_{proton}}{\rho \cdot Y_{step}}.
\]
For $\rho=10^5$\si{/cm^2} and $I_{protons} = 3 \times 10^5$ protons/spill, the $X_{step}$ = 3 cm. The target mover was driven at a constant speed during the flat top (FT, 4.8 s) in the 2016 run. However, we observed heterogeneity of beam density in the emulsion detector in the X direction (as shown in Figure \ref{positiondist} on the left). This was due to the time profile of the beam, which was not constant during the FT. The presence of such behaviour would be difficult to reproduce in simulations and thus becomes a source of error, particularly in the background estimation. Therefore, an effort to refine the control sequence was made. In the 2017 run, an additional scintillator counter (10 cm $\times$ 10 cm) was placed behind the target mover to monitor the intensity of the beam. The time profile measured by the scintillator is shown in Figure \ref{timeprofile}. The speed of the target mover was then configured every 0.2 s according to the actual intensity measured by the scintillator. The achieved position distribution in the emulsion detector is also shown in Figure \ref{positiondist} on the right, which has a flat distribution at an acceptable level. This measurement was done at the $I_{protons}=3\times 10^5$ protons/spill, corresponding to $X_{step}=3$ cm. 


The number of spills per module (12.5 \si{cm^2} $\times$ 10 \si{cm^2}) with the current achievement of 4 spills per line and 11 lines is 44 spills. The module exchanging time (stop beam, zone access and open beam) was measured as 20 min. Therefore, the time needed to expose one module would be about 1 h. The total effective beam time for future runs (368 modules) would then be 16 days.

\begin{figure}[htbp]
\begin{center}
\includegraphics[height=5.5cm]{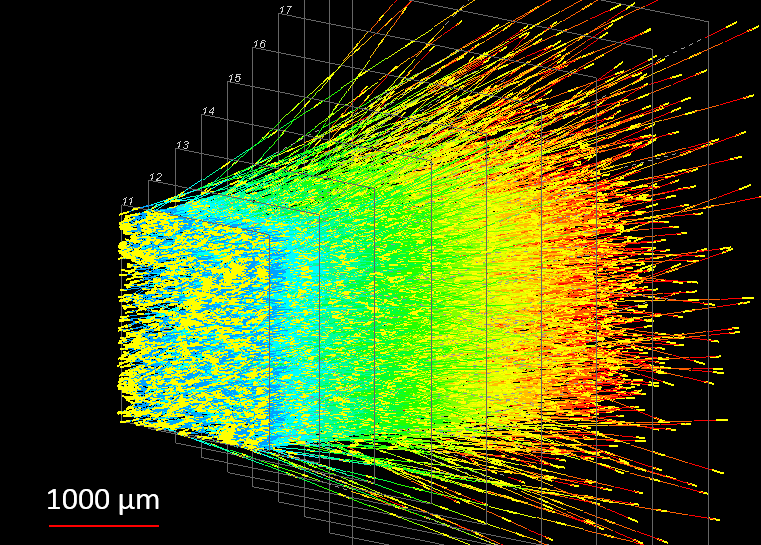}
\includegraphics[height=5.5cm]{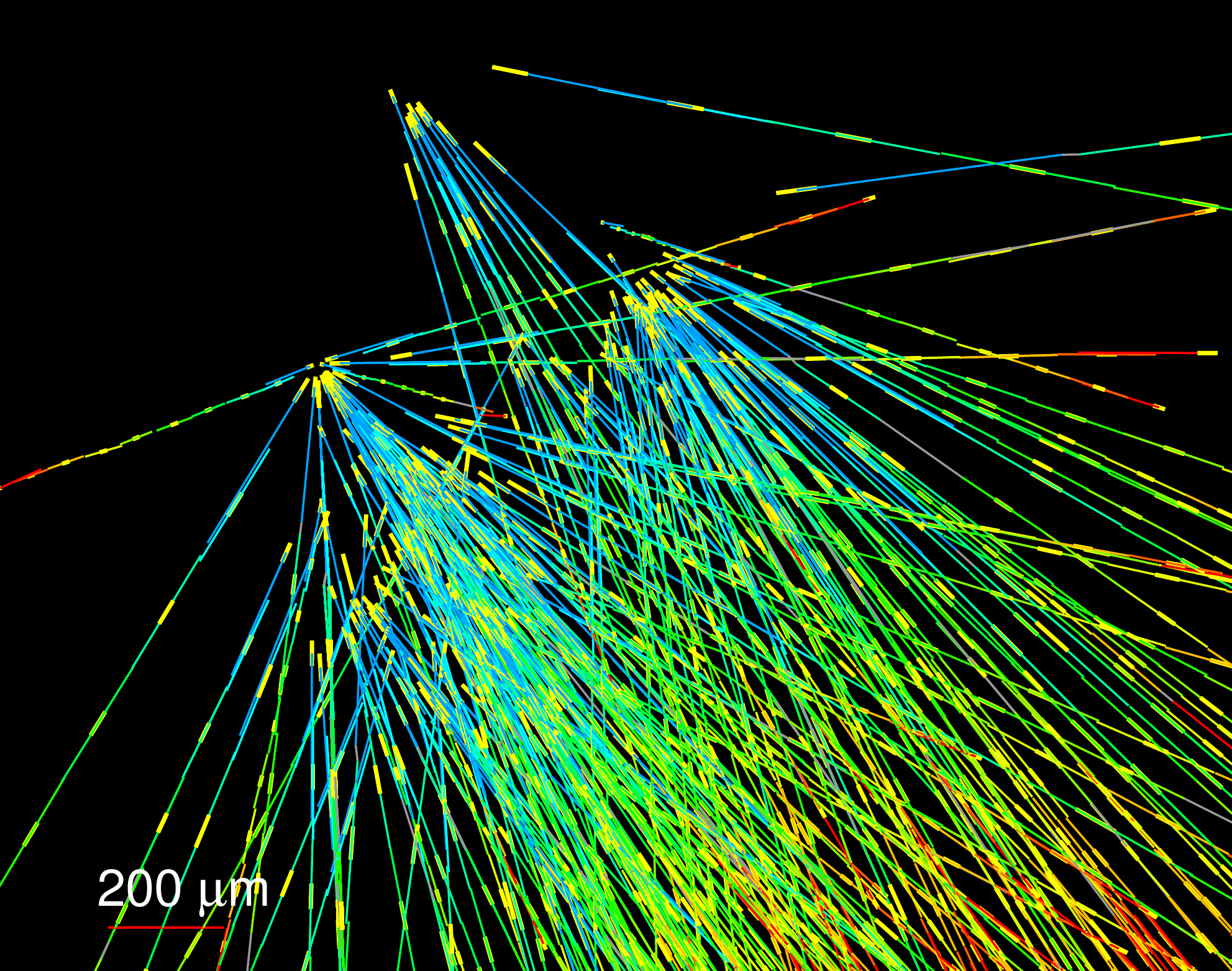}
\includegraphics[height=7cm]{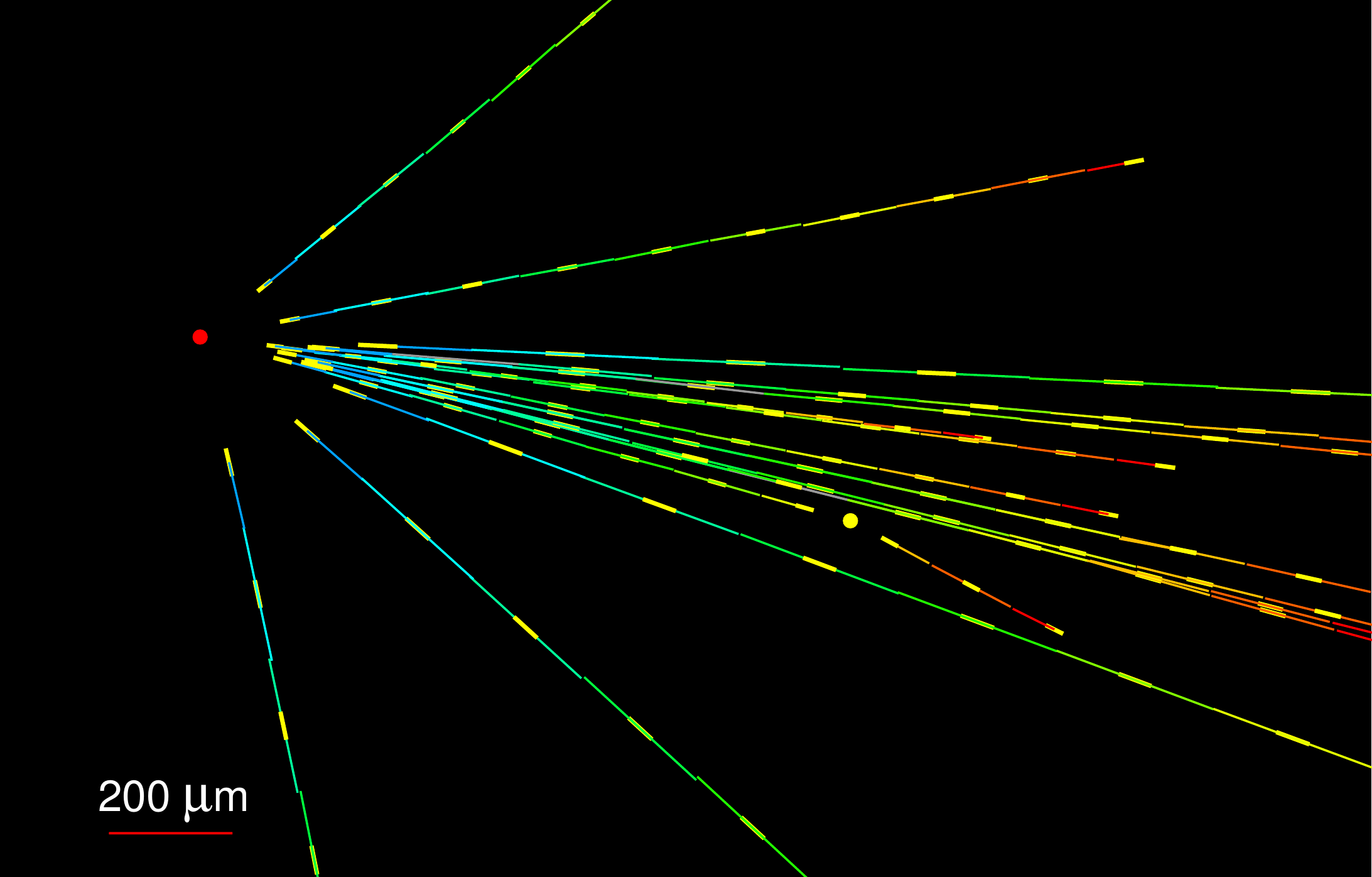}
\caption{Event display of reconstructed tracks (top left), proton interactions (top right) and an event with a possible decay topology (bottom). The primary and secondary vertices are indicated as the red and yellow points respectively. The flight length of the parent track is 3.73 mm, the kink angle is 66 mrad and the impact parameter of the kink daughter track with respect to the primary vertex is 256 $\mu$m. These are consistent with a charmed decay.
}
\label{testbeam_data}
\end{center}
\end{figure}

Ten modules were exposed to the 400-GeV proton beam, corresponding to 1/40th of the entire project.
We scanned the test beam samples with the HTS, which was able to read the track segments at the high track density of a few $10^5/\si{cm^2}$. A volume of 6 cm $\times$ 6 cm $\times$ 10 films was analysed initially. About 30,000 proton interactions were reconstructed and 
150 events with possible decay topology were detected (the number of expected charm events are about 30 events). The contamination of hadronic interaction background will be reduced by requiring the absence of nuclear fragments. Such detection has been performed by a dedicated scanning procedure \cite{anis,fukuda} in the OPERA experiment and that procedure is being adopted for this project. Figure \ref{testbeam_data} shows event displays of reconstructed tracks, proton interactions and an event with a possible decay topology found in the data. Further optimisation of the analysis scheme is underway. The angular reproducibility analysis shown before was also performed with the sample from this test beam. We also tested new films with thicker bases to improve the angular resolution. The completion of analysis is currently ongoing.

\section{Beam requirements and work schedule}

\begin{figure}[htbp]
\begin{center}
\includegraphics[width=\textwidth]{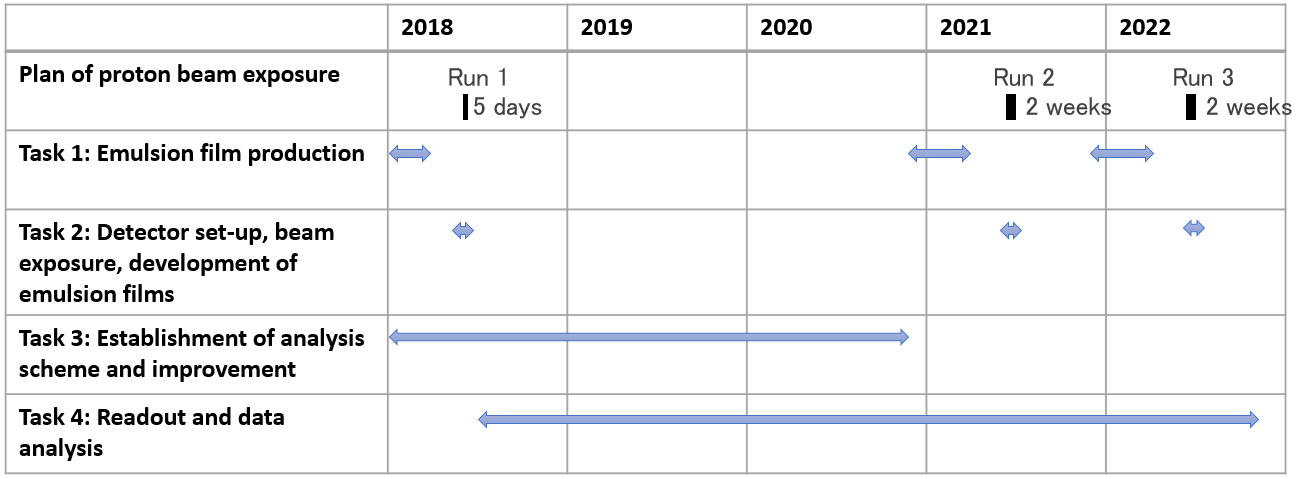}
\caption{Proposed work schedule for the project.
}
\label{schedule}
\end{center}
\end{figure}

The total effective beam time of 16 days would be split into three runs; one in 2018 and the other runs in subsequent years, depending on the schedule of CERN accelerators. This will enable us to optimise the experimental set-up in the second and third runs using the experience gained from the first. We would collect 1/10th of the required protons on target in the first run with five days of beam time (including the set-up time). We will analyse this first sample in the first three years, which would allow us to publish our research. The analysis of the second and third runs require less time since the analysis scheme will already be established. Figure \ref{schedule} shows the proposed work schedule for this project. 

\section{Cost}

We already have the facilities for emulsion handling and scanning. The major cost will be the emulsion detector ($\sim$250 kCHF). The fast emulsion scanning system is being upgraded (the cost of this is covered from existing funding sources for other research activity). Upgrading the existing scanning systems in Bern (and possibly other sites) with the high-precision read-out would cost $\sim$65 kCHF. 
The large amount of data will be processed with a GPU-based computing server (17 kCHF) and stored on a large disk server (100 TBytes; 25 kCHF). We are seeking for funding for both Japan and Europe. We have already secured funding for the 2018 run. We do not have any special request to CERN concerning infrastructure. 

\section{The collaboration}

The collaboration includes groups from five countries, Japan, Romania, Russia, Switzerland and Turkey. The collaborators are experienced scientists from the former DONUT, OPERA, and cosmic-ray and hadron experiments. Some are also members of the SHiP collaboration. Our expertise in the field of neutrino and hadron physics and emulsion detector technology provides a solid basis for the success of this project.

\end{document}